\documentclass[twocolumn,showpacs,preprintnumbers,amsmath,amssymb]{revtex4-1}
\pdfoutput=1 

\usepackage{graphicx}% Include figure files
\usepackage{dcolumn}% Align table columns on decimal point
\usepackage{bm}% bold math
\usepackage{upgreek}
\usepackage{array}
\usepackage{booktabs}
\usepackage{microtype}
\usepackage{cmap}
\usepackage{geometry}
\usepackage[dvipsnames]{xcolor}

\newcommand{\ME}[3]{\ensuremath{\left \langle \left. #1 \right.  \right| #2 \left| \left. #3 \right. \right \rangle}}

\usepackage{hyperref}
\hypersetup{colorlinks, citecolor={blue}, linkcolor={red}, urlcolor={violet}, pdftitle={Statistical Mechanical Treatments of Protein Amyloid Formation}, pdfauthor={John S. Schreck, Jian-Min Yuan}, pdfdisplaydoctitle}

\begin{document}
\title{Statistical Mechanical Treatments of Protein Amyloid Formation}% Force line breaks with \\

\author{John S. Schreck}

% \altaffiliation[Also at ]{Physics Department, Drexel University.}%Lines break automatically or can be forced with \\
\author{Jian-Min Yuan}%
\email{yuanjm@drexel.edu}
%\affiliation{%

\affiliation{Department of Physics, Drexel University, Philadelphia, PA 19104}%

\date{\today}% It is always \today, today,

\begin{abstract}
Protein aggregation is an important field of investigation because it is closely related to the problem of neurodegenerative diseases, to the development of biomaterials, and to the growth of cellular structures such as cyto-skeleton. Self-aggregation of protein amyloids, for example, is a complicated process involving many species and levels of structures. This complexity, however, can be dealt with using statistical mechanical tools, such as free energies, partition functions, and transfer matrices. In this article, we review general strategies for studying protein aggregation using statistical mechanical approaches and show that canonical and grand canonical ensembles can be used in such approaches. The grand canonical approach is particularly convenient since competing pathways of assembly and dis-assembly can be considered simultaneously. Another advantage of using statistical mechanics is that numerically exact solutions can be obtained for all of the thermodynamic properties of fibrils, such as the amount of fibrils formed, as a function of initial protein concentration. Furthermore, statistical mechanics models can be used to fit experimental data when they are available for comparison.
\end{abstract}

\maketitle
\section{Introduction}

Protein aggregation is an active, multidisciplinary science, with researchers and practitioners working in broad disciplines, including biophysics, medicine, biomaterials, and pharmaceuticals. With diverse perspectives, it is
not surprising that the papers on protein aggregation differ widely in their emphasis and
methodologies: from fundamental research related to molecular mechanisms and
aggregation pathways to searching for biomarkers, drug targets, even to
imaging of plaques in the brain, dissolution of fibrils and amyloids {\em in vivo}, {\em etc}. The
present article is only concerned with the fundamental investigations into the aggregation
mechanisms of amyloid formation related to neurodegenerative disease. Many
review articles have been written on the approaches based on molecular dynamical
simulations~\cite{thirumalai2003, Ma2006, Li2008, Urbanc2010}, as well as kinetic studies~\cite{Pallitto2001, powers, Knowles2009}. In our examination, we will focus instead on the statistical mechanical approaches to equilibrium assembly processes and present some new results while summarizing past approaches. 

%Early applications of statistical mechanical methods to the studies of protein
%problems can best be represented by the treatment of helix-coil transitions in proteins by Zimm and
%Bragg in 1950�s~\cite{zimm_bragg, lifson_roig, Scheraga1970, bloomfield_statistical_1999, Scheraga2002, chen_helix}. They assumed that each peptide bond linking amino acid residues together can exist in two states: a helical or non-helical (coil or sheet) state, and characterized the linear chain of residues with a partition function, which is a sum of all possible combinations of states.
%

Early applications of statistical mechanical methods to the studies of protein
problems can best be represented by the treatment of helix-coil transitions in proteins by Zimm and
Bragg in the 1950�s~\cite{zimm_bragg, lifson_roig, Scheraga1970, bloomfield_statistical_1999, Scheraga2002, chen_helix}. They assumed that each peptide bond linking amino acid residues together can exist in two states: a helical or non-helical state, and characterized the linear chain of residues with a partition function, which is a sum of all possible combinations of states.

Analogous to a one-dimensional Ising model~\cite{baxter}, Zimm and Bragg expressed the partition function in
terms of transfer matrices and solved the problem analytically in the large polymerization
limit, and also for finite chains~\cite{zimm_bragg}. Over the years, researchers have extended the original Ising-type models to study  sheet-coil~\cite{wsme1,wsme3, mattice_matrix_1984, mattice_beta-sheet_1989, sun_and_doig, hong_statistical_2008, Schreck2010} and helix-sheet~\cite{mattice_suppression_1984, hong, Schreck2010} transitions in proteins, as well as helix-coil~\cite{dutch_2001}, sheet-coil~\cite{Dutch2006}, and helix-sheet-coil~\cite{hong,Schreck2011} transitions in protein aggregates. 

It is in a similar spirit that our statistical mechanical treatment of protein aggregation
has been developed~\cite{Schreck2010}, which is the main subject of this article. Our formalism of the
aggregation processes was stimulated by other statistical mechanical studies. These
works will be briefly reviewed in Section~3, along with conceptual developments of
statistical mechanical techniques beyond those used in the Zimm-Bragg model. In
Sections 4--6, canonical ensemble and grand canonical ensemble treatments
of the aggregation processes will be separately presented, which is followed by a
Conclusion section. In~Section 2, we first review some properties of amyloid proteins and aggregates. 

\section{Amyloid Aggregation}

%%%%%%%%%%%%%%%%%%%%%%%%%%%%%%%%%%%%%%%%
	\begin{figure*}%f2
	\centering
	\includegraphics[width = 0.8\linewidth ]{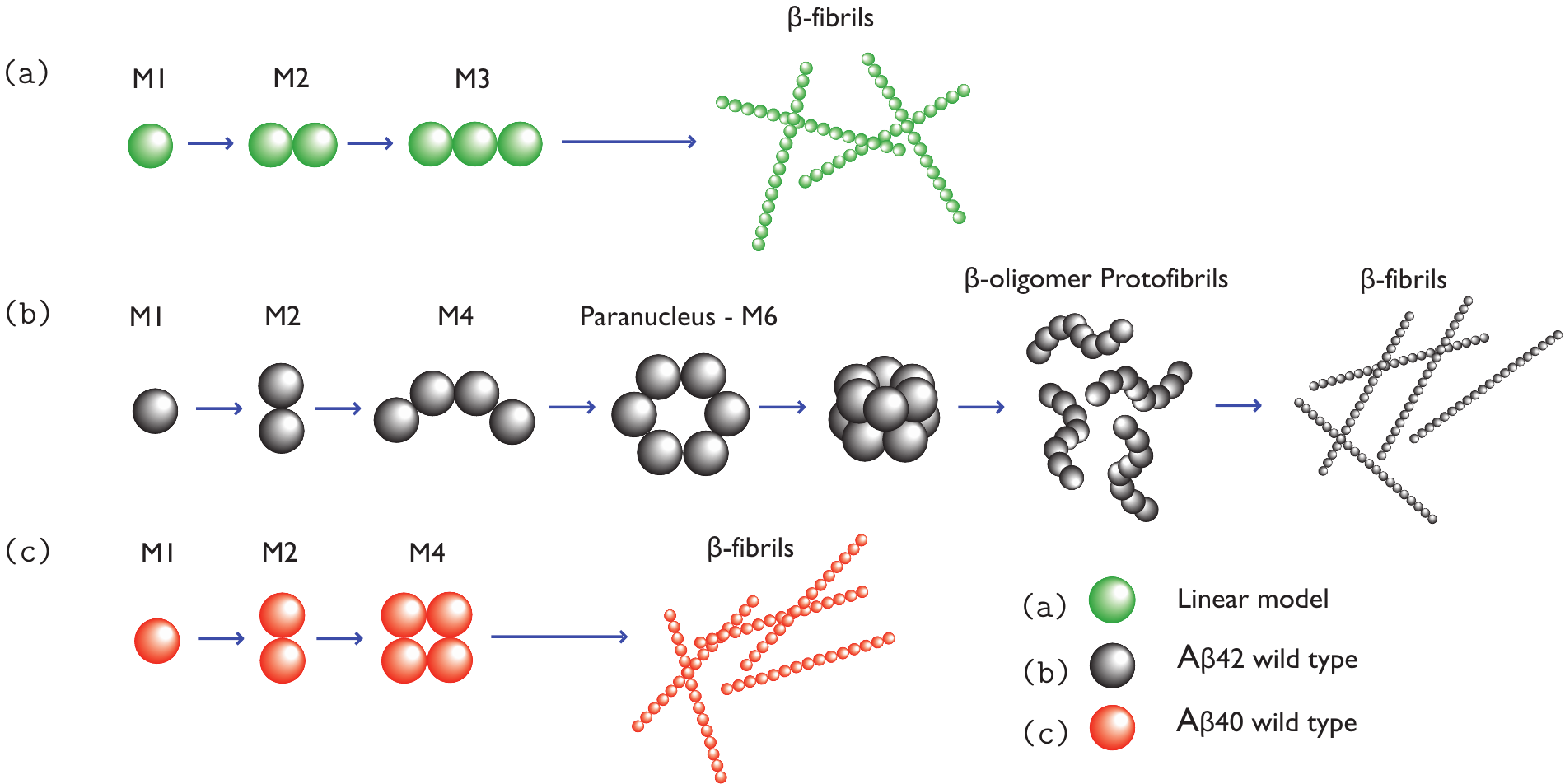}
	\caption{({\bf a}) A linear model for aggregation where $M$-mers grow one monomer at a time in a one-dimensional fashion; ({\bf b}) A model for A$\beta$(1-42) wild-type proteins aggregating into fibrils. Monomers assemble until reaching a critical concentration, for example, a hexamer, as illustrated. Hexamers then aggregate into more complex structures such as protofibrils, and eventually full fibrils; and ({\bf c}) a model for A$\beta$(1-40) wild-type protein aggregation where monomers join to form dimers, which may grow into protofibrils and eventually fibrils.}
	\label{assemb_pathway}
\vspace{-12pt}
	\end{figure*}
%%%%%%%%%%%%%%%%%%%%%%%%%%%%%%%%%%%%%%%%

To see the complication of protein aggregation processes, we use $\beta$-amyloid as an example. Under proper conditions, such as higher concentration, amyloid monomers can aggregate into dimers, trimers, tetramers, \dots, oligomers. These co-existing oligomers are in rapid kinetic equilibrium, making it difficult to determine their structures or numbers~\cite{Bitan2003}. Oligomers of the same size may exist in different conformations: some are partially ordered and some are disordered. As soluble oligomers grow larger they can become richer in $\beta$-sheet structures, but overall they lack secondary structures. They may resemble micelles, which have a spherical or cylindrical shape~\cite{Yong2002}. A$\beta$ oligomers seem to range in size, with a diameter ranging from 5--15 nm and molar mass ranging from 20--50 kDa up to 1 MDa~\cite{Lambert1998, Huang2000}.  Additionally, oligomers with ordered $\beta$-sheet or $\beta$-hairpin structures are believed to form protofibrils at a higher rate than their disordered counterparts~\cite{Li2008, goldschmidt2010, straub2011, morriss2012}. For example,  Li {\em et al.}~\cite{Li2008} showed in numerical simulations that a native chain has to unfold partially into an intermediate with beta-hairpin structure before ordered assembly can be formed. Since protein aggregation is thought to be a nucleation process, some of the ordered oligomers may act as paranuclei~\cite{Roychaudhuri2009}. Once formed, paranuclei can lead to the formation of protofibrils in down-hill fashion. The nucleation is illustrated in Figure~\ref{assemb_pathway}b,c. Since A$\beta$ oligomers may contain $\beta$-sheet structure, they could be pathway intermediate~\cite{Yong2002}. However, the same cannot be said about other oligomers, such as those comprised of prion proteins~\cite{Baskakov2002}. Additionally, A$\beta$(1-40) and A$\beta$(1-42) monomers may self-associate to form off-pathway globular assemblies including amylospheroids~\cite{Haass2007, Roychaudhuri2009} and $\beta$-amyloid balls~\cite{Hoshi2003} [formed by A$\beta$(1-40) only]. Both of these structures can grow to be quite large. 

\begin{figure*}
	\begin{center} 
	\includegraphics[width = 400 pt ]{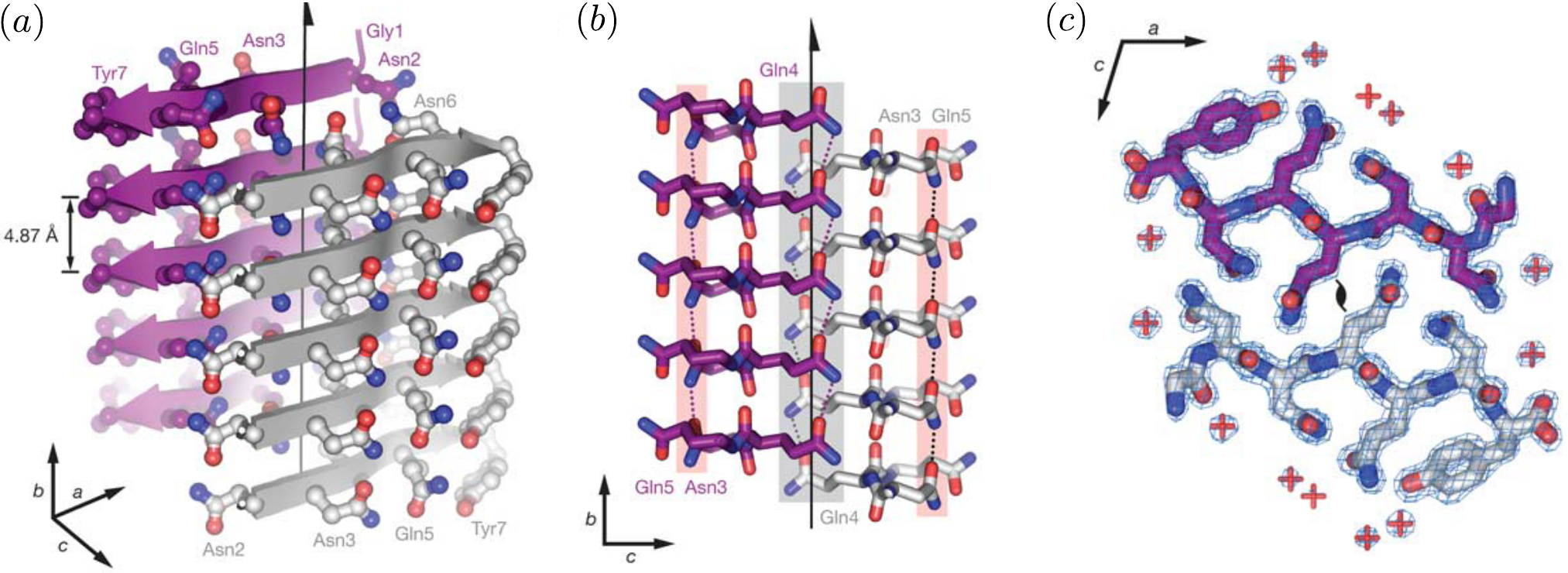}
	\end{center}
	\caption{In both ({\bf a}) and ({\bf b}), the illustrated cross-$\beta$ structure is the sequence segment GNNQQNY from the prion Sup35. Carbon atoms are purple or grey/white, oxygen atoms are red, and nitrogen atoms are blue.  In ({\bf a}), cross-$\beta$ structure is illustrated. Grey arrows represent the back-bone of a $\beta$-strand, and the side-chains are shown projecting from the strands. Purple arrows represent the strands residing in the back of the structure. The regions between the strands are referred to as the dry interfaces, whereas just outside of strands are wet interfaces. The fibril axis is indicated by an arrow running through the dry regions between the strands; ({\bf b}) Side view of the fibril. The H-bonds are formed between red carboxyl groups and blue amide groups from adjacent layers; in ({\bf c}), a top view of the fibrils shows the interdigitation of two $\beta$-sheets, referred to as the steric zipper. Within the steric zipper, water molecules are absent (a red plus sign indicates water). Both images are reprinted from Nelson {\em et al.}~\cite{Nelson2005, Eisenberg2012}.}
	\label{fibs}
	\end{figure*}

\begin{figure*}
	\centering
	\includegraphics[width = 400 pt ]{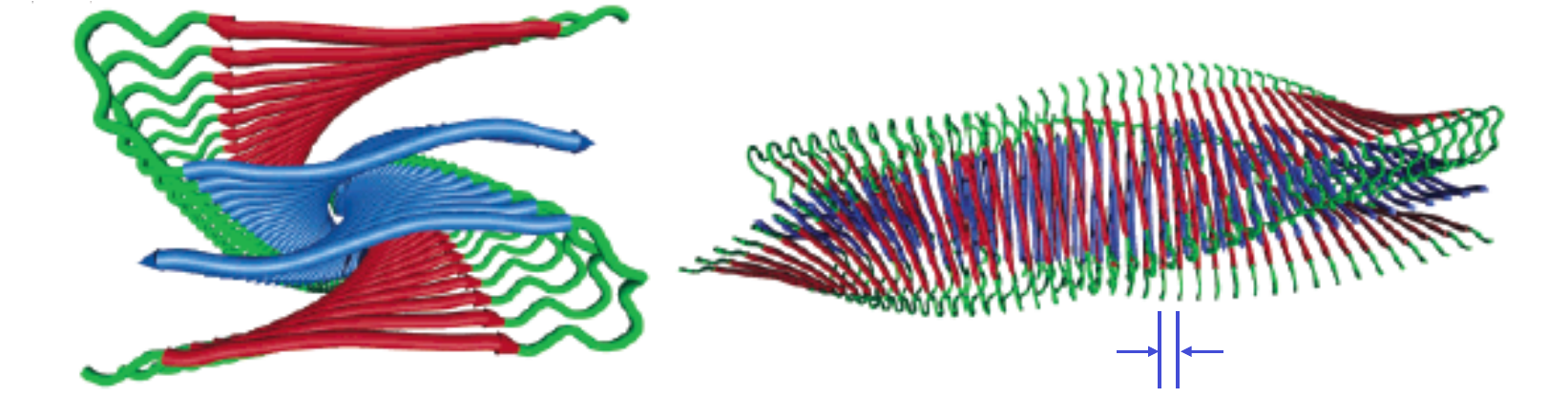}
	\caption{Cartoon illustrations of an A$\beta$ protofibril are shown. {\bf Left}: Looking down the axis of the fibril ({\em z}-axis); {\bf Right}: A sideview of the protofibril illustrating the twisted, helical pattern. Proteins are spaced at $\approx$5\,\AA, and the chiral twist of 0.833 degree/\AA ~was arbitrarily chosen for illustrative purposes. Image reprinted from Petkova {\em et al.}~\cite{Petkova2006}.  }
	\label{fibab}
\vspace{-12pt}
	\end{figure*}

Protofibrillar intermediates are heterogeneous, metastable aggregates already containing $\beta$-sheet regions in the core~\cite{Kheterpal2006}, but retaining some features that are similar to oligomers. The term ``protofibril'' has varying definitions throughout the literature, where for A$\beta$ the term could refer to structures ranging from 4--11 nm in diameter, up to 200 nm in length, and possibly even longer~\cite{Hartley1999, Harper1999}. Protofibrils are considered to be on-pathway during fibrillogenisis, and they could grow larger via monomer addition or merging with other oligomers or protofibrils~\cite{Hong2013, Schreck2013}. As protofibrils grow longer, the $\beta$-sheet region grows larger. Eventually, a stable and tight $\beta$-sheet network is formed in the core by the backbone H-bonds and the hydrophobic interactions of side chains~\cite{Kheterpal2006}. Figure~\ref{fibs} illustrates the cross-beta structure of fibrils comprised of the prion Sup35. These features are generally associated with protofibrils and fibrils~\cite{Kheterpal2006}. Another early intermediate thought to play a role in the formation of fibrils is the A$\beta$ protofilament, which can range in diameter from 2.5 nm up to about 6 nm~\cite{Serpell2000}, and are 50--100 nm long~\cite{Lashuel1999}. Several protofilaments may then merge and form fibrils, which may exhibit a twisted, helical ribbon structure~\cite{Makin2005} and contain highly-ordered $\beta$-sheet regions. A cartoon illustration of an A$\beta$ protofilament is shown in Figure~\ref{fibab}. The fibrils may even be composed of several segments with distinct morphologies and varying levels of ordered structure~\cite{brigita_private}. Additionally, mature fibrils have a diameter ranging from 7--12 nm, and may grower longer than 1 $\upmu$m~\cite{Makin2005}. Typically, fibrils are linear, non-branching structures. They contain very large amounts of $\beta$-structure, and are generally insoluble. Fibrils can further assemble into bundles~\cite{Ward2000}, and may form plaques outside the neurons. The total assembly process from monomers to fibrils for a simple 1D model is illustrated in Figure~\ref{assemb_pathway}a, and models for A$\beta$(1-42) and A$\beta$(1-40) fibrils are illustrated in Figures~\ref{assemb_pathway}b and \ref{assemb_pathway}c, respectively.

%%%%%%%%%%%%%%%%%%%%%%%%%%%%%%%%%%%%%%%%
%%%%%%%%%%%%%%%%%%%%%%%%%%%%%%%%%%%%%%%%

\section{ Statistical Mechanical Approaches to Protein Folding and Aggregation}

One of the earlier applications of statistical mechanical methods to protein systems is the Zimm-Bragg model~\cite{zimm_bragg, lifson_roig, Scheraga1970, Scheraga2002} originally developed for the studies of helix-coil transitions in proteins. Although the problems of macromolecular self-assembly that we are dealing with are quite different from conformational changes in proteins, the lesson that can learn from the model is fundamental. The Zimm-Bragg model is like an Ising model, which catches some essential features of a problem, and is solved rigorously using statistical mechanical methods~\cite{baxter}. Extending the ZB model to protein aggregation was first advanced by Oosawa and Kasai~\cite{oosawa}, Terzi {\em et al.}~\cite{terzi}, and more recently applied by van Gestel and de Leeuw~\cite{Dutch2006}, Schmit {\em et al.}~\cite{Schmit2011}, and others~\cite{lee, nyrkova, vander, kunes, nicodemi, badasyan, zamparo} to the study of macromolecular aggregation. These simple statistical mechanical models may be used to predict the average lengths of protofibrils and fibrils, and the fraction of protein molecules that assume various conformational secondary structures, including sheet, coil, and possibly helix. The main models summarized here focus primarily on computing partition functions for protofibrils and fibrils using simple effective Hamiltonians and transfer matrices. 
                            
%%%%%%%%%%%%%%%%%%%%%%%%%%%%%%%%%%%%%%%%
%%%%%%%%%%%%%%%%%%%%%%%%%%%%%%%%%%%%%%%%  
  
\subsection{Partition Function for Helix-Coil Transitions in Proteins}

Although protein aggregation is the subject of the article, we use the simplest model, {\em i.e.}, the  Zimm-Bragg model for helix-coil transitions in proteins, to illustrate the power of the statistical mechanical techniques. We first define the coil conformation of a protein residue as the reference state, which means its statistical weight is one~\cite{schellman1958, qian1992}. The parameter $s$ is the equilibrium constant for a coil residue converting into a helical residue, and relates to the free energy change of adding a helical residue to one end of a helical block, $\Delta G_s = -R T \ln s$, where $R$ is the gas constant. The nucleation step is to convert a coil residue into a helical residue in a chain of coil residues. The equilibrium constant associated with this event is $\sigma s$, where $\Delta G_{nuc} = -R T \ln \sigma$ is the additional free energy barrier the coil must overcome before converting into the first helical residue that can eventually be part of the helical block. Conventionally, $\sigma$ and $s$ are referred to as the {\it initiation} and {\it propagation} parameters. A single helical block in a protein thus has free energy $\Delta G = \Delta G_{nuc} + n_{h} \Delta G_{s} $, where $n_{h}$ is the number of helical residues in the block. 

In direct combinatorial approaches to solving the partition function, $Z_N$ for the helix-coil transitions in proteins of length $N$ amino acids is often expressed in terms of the initiation and propagation parameters. Some approximations can be made to simplify the mathematical expression for $Z_N$~\cite{qian1992, qian}. For example, the simplest model for helix-coil transitions assumes a single helical stretch where {\it all} of the residues in the protein are locked into the helix conformation. A more general approach assumes that the conformation of residues could depend on its neighbors and the nucleation and propagation of the chain occurs via a ``zipper'' mechanism, that is, one single stretch of helix can form along the chain of $N$ residues but may vary in length from one up to $N$. The partition function for the chain of $N$ residues in the zipper model can be written as: 
\vspace{-6pt}
\begin{eqnarray}
Z_N &=& 1 + \sum_{k=1}^{N} (N-k+1) \sigma s^k \nonumber \\ 
&=& 1 + \sigma s \frac{N-s-Ns}{(s-1)^2} 
\end{eqnarray}
where the term $(N-k+1)$ is the degeneracy in the number of ways putting $k$ helical residues next to each other along a chain of length $N$. 

The Zimm-Bragg (ZB) model for helix-coil transitions in proteins assumes that any number of helical stretches may form along the chain, where residues could be involved in short-ranged interactions with other residues. Each residue could assume either a coil or a helical conformation, thus for nearest-neighbor interactions between residues, there are four possible combinations of a pair of residues. That is, two residues at positions $j-1$, and $j$, respectively, could have states {\it cc}, {\it hc}, {\it ch}, or {\it hh}. The $j$th residue involved in the pair is assigned a weight depending on its conformation, and the conformation of the residue at $j-1$. For $cc$, the weight is one. If two neighboring residues adopt the $ch$ conformations,  the helical residue is assigned the weight $\sigma s$, which corresponds to the nucleation of a helical block.  On the other hand, in the original ZB model, the state $hc$ simply has weight one. That is, the growth of a helical block only proceeds in one direction along the chain. Finally, the $hh$ state corresponds with a weight of $s$ at the $j$th residue. The ZB model is summarized in Table~\ref{zbweights_table}.

\begin{table}
\centering
\begin{tabular}{ccc}
  \hline
    {\bf{\boldmath $j-1$}} & {\bf{\boldmath $j$}} & {\bf Weight }\\ 
  \hline 
  c & c & 1 \\ 
  h & c & 1 \\ 
  c & h & $\sigma s$ \\ 
  h & h & $s$ \\
  \hline
  \end{tabular} 
\caption{Summary of the ZB weights for two residues that are adjacent to each other in a~protein.}
\label{zbweights_table}
\end{table}

The partition function for the ZB model, $Z_N$, can be easily computed by using a transfer  matrix~\cite{kramers, onsager}. A transfer matrix is a device that can be used when a system of $N$ units can be decomposed into a subsystem of nearest neighbor, or next nearest neighbor, {\em etc}., interactions between all units. Since the residue located at a position $j$ along the chain only depends on the conformation of the residue at chain position $j-1$, the transfer matrix factors the partition function for some given energy function $E(r)$ of a system of $N$ units as: \\

\begin{eqnarray}
\label{partition_function}
Z_{N} &=& \sum_{r} e^{-\beta H(r)} \nonumber \\
&=& \ME{f}{T^{N}}{i} \nonumber \\
&=& \sum_{i=1}^{N_{\lambda}} c_i \lambda_i 
\end{eqnarray} 
%\left \langle \left. #1 \right.  \right| #2 \left| \left. #3 \right. \right \rangle
where the vectors $\left \langle \left. f \right.  \right| $ and $|\left. i \right. \rangle$ represent the states of the residues at either end of the chain, respectively, $\lambda_i$ and $N_{\lambda}$ are the eigenvalues, and total number of eigenvalues of $T$, respectively. Additionally, in the third line of Equation~\ref{partition_function}, the coefficients $c_i$ take into account the effect of the boundary conditions. The notation $r = (r_1, r_2, ..., r_N)$ refers to the set of all $N$ spins that represent the states of the residues, where $r_i$ denotes the state of the $i$th residue. 

%\begin{figure}
%\begin{center}   
%\includegraphics[width=350 pt ]{zbweights}
%In (a), the ZB model for helix-coil transitions in proteins is illustrated. Circles represent amino acid resides while squares are peptide groups. In $\alpha$-helices, H-bonds connect the ith and i+4 residues along the chain when the residues at i+1, i+2, and i+peptide units receive the coding and weight, but residues are often referred to as being either ``helix" or ``coil". Once a residue is locked into a conformation, the peptide unit immediately to it's left along the chain picks up the weight. In (b), a simple model for protein aggregation is illustrated. Note that the conformation of the monomer, or proteins in aggregates, is not taken into account.
%\caption{ The ZB model for protein aggregation is illustrated, where the proteins in aggregates can be sheet (white circles), or coil (black circles) in conformation. The weight associated with each protein is listed. }
%\label{zbweights}
%\end{center}
%\end{figure}

The transfer matrix elements represent the probability that a residue occupies a different state from its neighbor. Thus, the transfer matrix used in the ZB model for helix-coil transitions in proteins has the following form:
\begin{eqnarray}
\label{zbtransfer}
T_{i} &=&\left( \begin{tabular}{c|cc}
 & ~h~ & c~ \\
\hline
h~ & ~s~ & 1~ \\ 
c~ & ~$\sigma$s~ & 1~\\
\end{tabular} \right)
\end{eqnarray}
where the column $\left(h,c\right)$ is the $(j-1)th$ state, and the row $\left(h,c\right)$ the $j$th state. As indicated in  Equation~(\ref{partition_function}), the matrix T can be readily diagonalized and the partition function written in terms of its eigenvalues. We~show the diagonalization as an example in the next section while discussing the helix-coil and sheet-coil transitions in equilibrium protein aggregation. For periodic boundary conditions, the $c_i$'s all equal unity in Equation~(\ref{partition_function}) and the partition function can be written as:
\begin{eqnarray}
\label{tranfff}
 Z &=& \operatorname{tr}\left(T^N\right) \nonumber \\
 \label{blahblah}
 &\approx& \lambda_1^N
 \end{eqnarray}
 where ``{\em tr}'' refers to the trace operation and $\lambda_1$ is the largest eigenvalue of Equation~(\ref{zbtransfer}). Equation~(\ref{blahblah}) is valid only when $N$ is large and becomes exact in the thermodynamic limit when $N \to\infty$. The partition function then reduces to:
 \begin{equation}
 \label{logz}
Z = \left( \frac{1+s+\sqrt{(s-1)^2+4 s \sigma } }{2} \right)^N
  \end{equation} 

\subsection{Thermodynamic Properties of Proteins}

Once the partition function for the protein is known explicitly using the ZB theory, some average quantities can be defined and compared with experiments. The average fraction of residues in a chain of length $N$ that are helical is referred to as the helicity, $\theta$. It can be defined as:
\begin{eqnarray}
\label{theta}
\theta &\equiv& \frac{1}{N_H} \frac{\partial~\text{ln}~Z}{\partial~\text{ln}~s} 
\end{eqnarray}
\vspace{6 pt}
where $N_H$ is the maximum number of helical residues in the chain. The helicity is akin to measuring the magnetization of spin systems when an external magnetic field is supplied. For the helical protein, the average number of helical segments, $v$, and the average helical length, $L$, are also found  using Equation (\ref{partition_function}),  

\begin{eqnarray}
v &=&\frac{\partial~\text{ln}~Z}{\partial~\text{ln}~\sigma} \\
L &=& N_H \frac{\theta}{v} = \frac{s}{\sigma}\frac{\partial Z/\partial s}{\partial Z/\partial \sigma} 
\label{lengthzb}
\end{eqnarray}

Similar averages are calculated for the sheet-coil and helix-sheet-coil systems. In general for a chain of length $N$, the average number of residues having the property $x_{j}$ is given by:
\begin{equation} 
\label{general_thermo}
\langle n_{j} \rangle =\frac{1}{N_{H}} \frac{\partial~\text{ln}~Z}{\partial~\text{ln}~x_{j}}
\end{equation}
where $j$ could refer to helix, coil, sheet, {\em etc}. Thus, the helicity and the average length of helical segments for long chains are now easily computed by inserting Equation~(\ref{logz}) into Equations~(\ref{theta}) and (\ref{lengthzb}), we are  left with:
\begin{eqnarray}
\label{theta_param}
\theta &=& \frac{1}{2}+\frac{s-1}{2 \sqrt{(s-1)^{2}+4 \sigma s}}  \\
L &=&1+\frac{2 s}{1-s+\sqrt{(1-s)^{2}+4 \sigma s}}
\end{eqnarray}
and a similar result can be derived for $\theta$ and $L$ when using finite boundary conditions. 

\section{Equilibrium Protein Aggregation }

%In 1961 Oosawa and Kasai constructed a model for equilibrium protein aggregation using ideas from the helix-coil theory that were being developed at the same time. First, in the model the total number of proteins in the system is fixed and is denoted by $m_{tot}$. Next, the model assumes that a dimer is the smallest aggregate that may form. The chemical reaction for dimer formation represents the nucleation of an aggregate~\cite{oosawa, Oosawa1975, terzi} and can be quantified by an equilibrium constant denoted $K_{eq}$: 
%\begin{equation}
%A + A \overset{ K_{eq} }{\leftrightharpoons} A_2
%\end{equation}
%where A represents monomer, $A_k$ represents the aggregate containing $k$ proteins and is referred to as  a k-mer. 

In 1961, Oosawa and Kasai constructed a model for equilibrium protein aggregation using ideas from the helix-coil theory that were being developed at the same time. First, in the model the total number of proteins in the system is fixed and is denoted by $m_{tot}$. Next, the model assumes that a dimer is the smallest aggregate that may form. The chemical reaction for dimer formation represents the nucleation of an aggregate~\cite{oosawa, Oosawa1975, terzi} and can be quantified by an equilibrium constant denoted $K_{eq}$: 
\begin{equation}
A + A \overset{ K_{eq} }{\leftrightharpoons} A_2
\end{equation}
where {\em A} represents monomer, $A_k$ represents the aggregate containing $k$ proteins and is referred to as  a {\em k}-mer. 

The nucleation equilibrium constant is often denoted $K_{eq}=\sigma s$, for example, in Terzi's model~\cite{terzi}. The concentrations for monomers, $n_1$, and dimers, $n_2$, can also be written as: 

\begin{equation}
n_2 = \sigma s n_1^2
\end{equation}

Once a dimer is formed, then trimer, $\dots$, {\em k}-mer may form by successive addition of a protein to an aggregate. Any of these reactions can be described by the monomer addition mechanism,  represented by:
\begin{equation}
A_i + A \overset{ s }{\leftrightharpoons} A_{i+1}
\end{equation}
where $s$ is the equilibrium constant. Thus, if the equilibrium constant for monomer addition is $s$, we can write the equilibrium concentration for the $k$-mer as: 

\begin{equation}
n_k = \sigma s^{k-1} n_1^k
\end{equation}

Since the total protein mass in the system is conserved, $m_{tot}$ can be written in terms of the concentrations of monomers and aggregates as: 

\begin{eqnarray}
\label{con_mass}
m_{tot} &=& \sum_{k=1}^{\infty} k n_{k}\nonumber \\
&=& 1 \cdot n_1 + 2 \cdot n_2 + \dots + k \cdot n_k + \dots \nonumber \\
&=& 1 \cdot n_1 + 2 \sigma s n_1^2 + \dots + k \sigma s^{k-1} n_1^k + \dots
\end{eqnarray}

Therefore, in the thermodynamic limit $N \to\infty$ the expression can be written as: 
\begin{eqnarray}
\label{zipper_agg}
m_{tot} &=& n_1 \left(1 + \sum_{k=2}^{\infty} k \sigma s^{k-1} n_1^{k-1} \right) \nonumber \\
&=& n_1 \left( 1- \sigma + \frac{\sigma}{(1-s n_1)^2} \right) 
\end{eqnarray}
where the sum converges when $s n_1 < 1$. If $m_{tot}$ is known, Equation~(\ref{zipper_agg}) can be solved for the monomer concentration $n_1$. 

\subsection{A Generalized Zimm-Bragg Model for Protein Aggregation}

A more recent approach to equilibrium peptide assembly introduced by van Gestel and  van der Schoot relates the concentrations of protein aggregates to equilibrium partition  functions~\cite{Dutch2006}. The partition functions of aggregates are expressed in terms of ZB  initiation and propagation parameters and a transfer matrix. The model describes 1D  protein aggregation, where the protein monomer is dominated by coil, sheet, or helical  conformations, discussed below, and may participate in short-ranged interactions with other proteins. Hence, the aggregates may exhibit various degrees of conformational order, where  helix-coil~\cite{dutch_2001, dutch_2003}, sheet-coil~\cite{Dutch2006, dutch_2008}, or helix-sheet transitions~\cite{Schreck2010, Schreck2011} may occur. 

This modeling approach is consistent with a recent set of experiments~\cite{Sarroukh2011}, for example, that have shown that oligomers of A$\beta$(1-40) and A$\beta$(1-42) are dominated by antiparallel $\beta$-sheet structures, while their fibrils are mainly characterized by parallel $\beta$-sheet structures. Thus major conformational changes may take place somewhere between the oligomer and the fibril formations, and using Ising-like ZB models may be advantageous for studying conformational transitions involved in protein aggregation. 

The isolated monomer in the ZB model for aggregation is assumed to be a natively unstructured protein. A ``helix'' protein is defined if $\theta_{\text{helix}} >\theta_{\text{sheet}}$ and $\theta_\text{helix} >\theta_\text{coil}$, where $\theta_\text{helix}$ can be defined by using Equation~(\ref{theta_param}), or a related equation for sheets. Similar definitions define ``sheet'' and ``coil'' proteins. The random coil does not have stable secondary structures. This toy model is suitable for understanding how proteins form fibrils in 1D, however, a word of caution is that, as mentioned above, amyloid formation can be sequence-dependent, for example, A$\beta$(1-40) and A$\beta$(1-42), have {different pathways~\cite{Bitan2003}}.

All of the aggregate species in a system of volume $V$ are assumed to be in kinetic equilibrium with each other, where interest lies in studying the thermodynamic properties of these aggregates. The system of proteins and aggregates is also assumed to be well mixed, and containing a fixed amount of protein mass given by Equation~(\ref{con_mass}). If the system contains only low concentrations of proteins and aggregates, the solution properties of the aggregates can be calculated by employing a standard ideal gas approximation for the $k$-mers, where the partition function for the system can be written as: 
\begin{equation}
\label{ztot}
Z_{T} = \prod_{k} \frac{ Z_{k}^{n_{k}}} { n_{k} !}
\end{equation}
where $Z_{k}$ is the canonical partition function for the $k$-mer. The number distributions $n_{k}$ per unit volume define the number densities $\rho_{k} \equiv n_{k}/V$. The relative densities of $k$-mers can be derived by considering the total free energy density $\Delta \mathcal{F}$, which may be written compactly for a system containing $N$ number of proteins, as: 
\begin{eqnarray}
\label{free}
\Delta \mathcal{F} &=& \sum_{N=1}^{\infty} \rho(N) \left[ \ln \rho(N) - 1 - \ln Z(N) \right] 
\end{eqnarray}
which contains an entropy of mixing term as well as the free energy of the aggregate of size $N$. We can minimize Equation~(\ref{free}) with respect to the total number density $\rho_{T}$ and subject to constraint given in Equation~(\ref{con_mass}), {\em i.e.}, conservation of mass, 
%\begin{equation}
%\frac{ \partial}{\partial n_k} \left( \Delta \mathcal{F} - \mu \frac{N_T}{V} \right) = 0
%\end{equation}
which yields for the number densities:
\begin{eqnarray}
\label{dens}
\rho(N) &=& Z(N) \exp(\mu N)
\end{eqnarray}
where $\mu$ is the Lagrange multiplier, and is realized as the chemical potential of a protein, $\rho(N)$ is just the $N$th moment of quasi-grand ensemble $\Omega$ = $\prod_{k=1}^{N_T} \rho(k)$. As in the 1D model by van Gestel {\em et al.}~\cite{dutch_2001, Dutch2006}, the state of an aggregate is directly coupled to the aggregate size distribution.  

A generalized ZB model for protein aggregation can now be defined by an effective Hamiltonian. The~effective Hamiltonian is used to find a transfer matrix by assuming the interactions between aggregates are described by a nearest-neighbor, Ising-like model, in which the protein could be in any of the two states: a sheet (or helix) or coil conformation. The interactions include  the free energy $R<0$, which describes the inter-facial tension between adjacent sheet and coil proteins in an aggregate. The parameter $P>0$ represents the interaction between two neighboring proteins, where one of the proteins located at position $j$ along the chain is in a sheet conformation. $P$ for sheets is measured relative to the coil interaction energy, which was taken to be zero. Additionally, the free energy $K>0$ quantifies  a polymerizing interaction between any two monomers along the 1D lattice that does not depend on their respective conformations. The effective Hamiltonian used by van Gestel and others has the following form for $N$ monomers~\cite{grosberg1994, Dutch2006}:
\begin{equation}
\label{dutch_energy}
E(r) = - \frac{1}{2} R \sum_{i=1}^{N-1} \left( r_i r_{i+1} - 1 \right) + \frac{1}{2} P \sum_{i=1}^{N} \left( r_i + 1 \right) + (N-1) K
\end{equation}
where $r=(r_1, r_2, ..., r_N)$ and $r_i$ can take on values \{$1,-1$\} corresponding to the spin states \{$\uparrow, \downarrow$\} in the Ising model, and to \{$s, c$\} in a Zimm-Bragg model for sheet-coil aggregates. With periodic boundary conditions, the partition function for the two-state model can be written as: 
{\small \begin{eqnarray}
\label{tftftf}
Z_{N} &=& \sum_{r} e^{-\beta E(r)} \nonumber \\
&=& e^{(N-1)K} \sum_{r} \exp \left[ \frac{1}{2} R \sum_{i=1}^{N-1} \left( r_i r_{i+1} -1 \right)  - \frac{1}{2} P \sum_{i=1}^{N} \left( r_i + 1 \right)  \right] \nonumber \\
&=& k^{N-1} \sum_{r} T(r_{1},r_{2}) T(r_{2},r_{3}) \cdots T(r_{N-1},r_{N}) T(r_{N},r_{1}) 
\end{eqnarray}}
where $e^{\beta (N-1)K}\equiv k^{N-1}$ with $\beta=1/k_{B} T$, and the parameters are redefined as $K\equiv K/k_B T$,  $R\equiv R/k_B T$, $P\equiv P/k_B T$. The transfer matrix can be written as:
\begin{eqnarray}
\label{ising}
T(r, r') &=& \exp \left[- \frac{R}{2} (r r'-1) + \frac{P}{2}  (r'+1) \right] \nonumber \\
&=& \left( \begin{array}{cc}
1 & \sqrt{\sigma_{1}}  \\
s_{1} \sqrt{\sigma_{1}}  & s_{1}  
\end{array} \right)
\end{eqnarray}
where $\sigma_1 = \exp(-2R)$ and $s_1 = \exp( P)$ are the initiation and propagation parameters for the aggregate system. We note that Equation~(\ref{ising}) and Equation~(\ref{zbtransfer}) yield the same characteristic equation, hence they predict the same thermodynamics results. The difference in the formulation of the folding model versus the aggregation model is that the helical regions of proteins, for example, may only elongate in one  direction, while helical aggregates may grow longer at either side. While amyloid~fibrils are mainly dominated by beta structure, helical conformations may play some roles during fibril formation~\cite{marini2002left, takahashi2000mutational}. It may also be advantageous to study statistical mechanical models that can describe the interaction of molecules capable of binding to A$\beta$ structure in fibrils, which may inhibit oligomer formation in the early stages {of aggregation~\cite{takahashi2008peptide}}.

\section{Partition Functions for Fibrils} 

Amyloid formation is generally believed to be dominated by 1D or quasi-1D chains of proteins, which may then bundle into protofibrils and fibrils. It is because of this fact that the transfer matrix formulation in statistical mechanics, if extended successfully, is a powerful technique for the studies of amyloid formation. We focus on this extension in this and the following sections. 

\subsection{Potts Model for 1D Filaments}

To include helix, sheet, and coil conformations in a single model, a Potts model~\cite{wu} for 3-state proteins can be used~\cite{Schreck2011}. The spin variable, $r$, may now assume values of $r=0$ for coil proteins, $r=1$ for helical proteins, and $r=2$ for sheet proteins. Interactions between proteins are assumed to be with nearest-neighbors only so that a dimensionless Potts model for the aggregate containing $N$ number of proteins can be written as:  
\begin{equation}
\label{hamil_hcs}
\begin{split}
-\beta H_{fil} &= - \sum_{i=1}^{N-2} R(r_{i},r_{i+1}) \left[1-\delta(r_{i}, r_{i+1}) \right] \\ & \quad + \sum_{i=1}^{N-1} \left( P_{1} \delta(r_{i},1) +  P_{2} \delta(r_{i},2) \right) \end{split}
\end{equation}
where the Kronecker delta $\delta( x, y )$ equals one if $x=y$ and zero otherwise. Equation~(\ref{hamil_hcs}) is illustrated in Figure~\ref{zbweights}. Like in Equation~(\ref{dutch_energy}), the initiation parameters are defined as {$\sigma(r_{j},r_{j+1}) \equiv \exp(-2 R(r_{j},r_{j+1}))$ and $R(r_{j},r_{j+1})$ $>$ 0 is the free energy of the interfacial tension between proteins at positions $j, j+1$ that are not in the same conformation~\cite{dutch_2001}. Thus, three types of interfaces between neighboring proteins in a generalized model are possible: {\em hc} or {\em ch}, {$R(0,1)=R(1,0)\equiv R_{1}$; {\em sc} or {\em cs}, $R(0,2)=R(2,0)\equiv R_{2}$; and {\em sh} or {\em hs}  $R(1,2)=R(2,1)\equiv R_{3}$. The notation can be simplified by letting $\sigma(1,0)=\sigma(0,1)\equiv \sigma_{1}$, $\sigma(0,2)=\sigma(2,0)\equiv \sigma_{2}$, and finally $\sigma(2,1)=\sigma(1,2)\equiv \sigma_{3}$.

\begin{figure*}
\centering
\includegraphics[width=350 pt ]{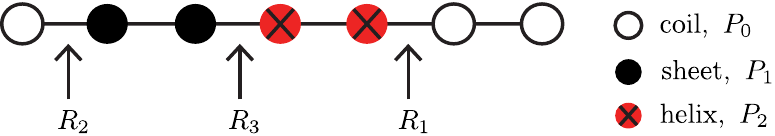}
\caption{ A ZB model for protein aggregation is illustrated, where the proteins (circles) in aggregates can be coil (white), sheet (black), or helix (red marked with X) in conformation. The free energies associated with each conformation are listed, as well as the interfacial free energies $R_1$, $R_2$, and $R_3$ between helix-coil, sheet-coil, or helix-sheet regions, respectively.  }
\label{zbweights}
\vspace{-12pt}
\end{figure*}
\vspace{12 pt}

The propagation parameters $s_1$ and $s_2$ are associated with the free energies $P_1$ and $P_2$ that refer to the interaction between the $i$th protein that is helix or sheet, respectively, and the nearest neighbor protein at location $i+1$. The coil protein interaction energy is assumed to be zero so that it may serve as a reference state. The transfer matrix can then be written as 
\begin{eqnarray}
\label{1dtransfer}
T(r, r') &=& \exp\left\{ - R(r, r') \left[ 1-\delta(r, r') \right] + P_{1} \delta(r, 1) + P_{2} \delta(r, 2) \right\} \nonumber \\ 
&=& \left( \begin{array}{ccc}
1 & \sqrt{\sigma_{1}} & \sqrt{\sigma_{2}}  \\
s_{1} \sqrt{\sigma_{1}}  & s_{1}  & s_{1} \sqrt{\sigma_{3}} \\ 
s_{2} \sqrt{\sigma_{2}}  & s_{2}  \sqrt{\sigma_{3}}  & s_{2} \\ 
\end{array} \right)
\end{eqnarray}
and the partition function for $N>2$ proteins in the aggregate can be calculated by diagonalizing Equation~(\ref{1dtransfer}) and plugging into Equation~(\ref{partition_function}). 

The Ising-like ZB model for sheet-coil (or helix-coil) transitions in aggregates, Equation~(\ref{dutch_energy}), can be recovered by writing the $q=2$ version of the effective Hamiltonian given in Equation~(\ref{hamil_hcs}). By choosing internal states such that $r_{i}=-1$ is the coil state and $r_{i}= +1$ is the helix state, and by making the substitution $\delta( r_{i}, r_{j} )$ = $\frac{1}{2} \left(1 + r_{i} r_{j} \right)$, the effective Hamiltonian and corresponding transfer matrix for 1D, two-state helix-coil is recovered~\cite{dutch_2001}. Next, the Hamiltonian for the 1D model given by Equation~(\ref{hamil_hcs}) can be applied to describe the interactions between proteins in aggregates on quasi-1D lattices. 

\subsection{Simple Model for Fibrils}

To describe the fibrils, which may contain several filaments that can be described by Equation~(\ref{hamil_hcs}), various models could be used. For example, two identical filaments could align in register, and all of the proteins could be in the sheet conformation already, or all coil, or some combination of sheet, coil, and even helix proteins. A simple Hamiltonian for the all-sheet case can be written as~\cite{nyrkova, Dutch2006, dutch_2008}:
\begin{equation}
- \beta H = L_y ( L_x - 1)(K + P_1) + L_x(L_y-1) B
\end{equation}
where the free energy $B>0$ describes a lateral binding interaction between two sheet proteins on different filaments, $L_y$ refers to the the number of filaments, and $L_x$ is the length of each filament. Additionally, in our approach $K$ was the polymerizing interaction between two adjacent proteins in a fibril, and $P_1$ was free energy of an interaction between a sheet protein and one of its neighbors.   Other effective Hamiltonians could also be written to describe the case where the filaments are not aligned in register with each other, as well as cases where the protein conformation may also play a role in the assembly of the filaments into full fibrils~\cite{nyrkova, dutch_2008}. 

\subsection{Quasi-1D Models for Aggregates}

In addition to conformational changes in filaments and fibrils, another complication is that the kinetic pathways to the formation of fibrils seem to be sequence-dependent~\cite{Bitan2003}.  The A$\beta$(1-40) isoform solution is abundant in dimers, then trimers, tetramers, $\dots$, in decreasing order.  However, the A$\beta$(1-42) isoform is more abundant in hexamers and pentamers than in dimers and trimers~\cite{Bitan2003, Urbanc2010}.  These facts seem to be consistent with recent experiments~\cite{Kheterpal2006, Shankar2008, Sarroukh2011}, which indicate that the A$\beta$(1-40) dimer is particularly stable and contributes to protofibril formation.  On the other hand, circular hexamers seem to play a role in the protofibril formation of A$\beta$(1-42)~\cite{Bitan2003, Urbanc2010}. 

We can model the equilibrium aggregates of A$\beta$(1-40) and A$\beta$(1-42) proteins by using finite strips of a two-dimensional $N \times N$ square lattice. In Figure~\ref{ab40} and Figure~\ref{cube}a, two identical 1D lattices stacked in-register, that is, a strip lattice of width two, are used to represent an A$\beta$(1-40) aggregate. For A$\beta$(1-40), we assume that the smallest equilibrium aggregate is the critical nucleus, which in this case is taken to be the dimer, while the smallest aggregate for A$\beta$(1-42) is the hexamer, as illustrated in Figure~\ref{ab42fig} for a strip lattice of width six.

%%%%%%%%%%%%%%%%%%%%%%%%%%%%%%%%%%%%%%%%
	\begin{figure}
	\centering
	\includegraphics[width = 300 pt  ]{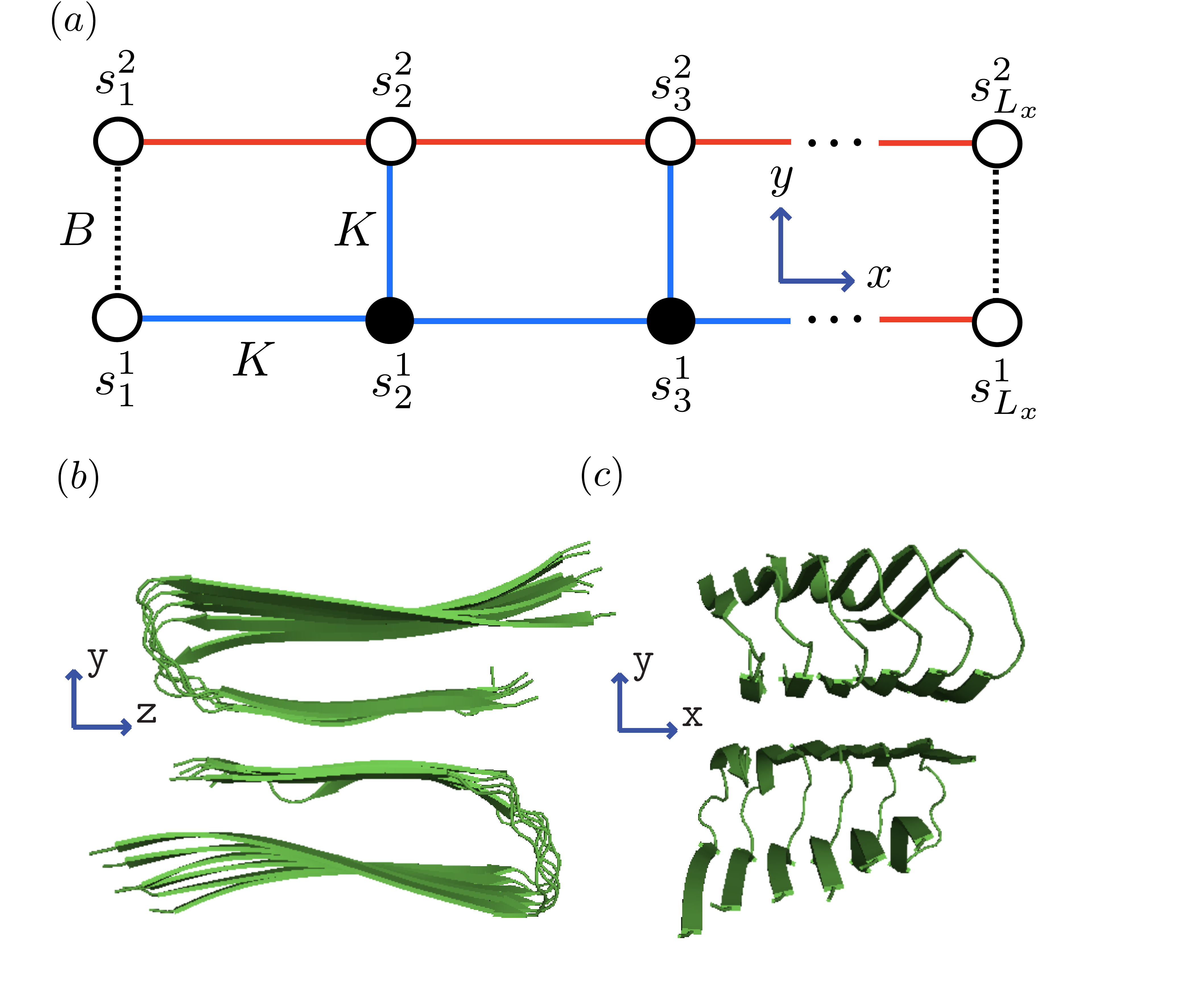}
	\caption{({\bf a}) Front-view ({\em y}-{\em z} plane) of a strip lattice representing an aggregate of \protect A$\beta$(1-40) proteins. Black circles corresponds to coil proteins while white circles denote sheet or helix proteins. Dashed lines represent interactions between proteins in the {\em y}-direction whereas solid lines are the interactions between proteins along the {\em x}-axis; ({\bf b})~Side-view ({\em x}-{\em y} plane) of A$\beta$-40 proteins illustrating the steric zipper; and ({\bf c}) strip lattice representation of the A$\beta$-40proteins, where the parameters $B$ and $K$ are illustrated.}
	\label{ab40}
\vspace{-12pt}
	\end{figure}

The position of a vertex within the strip is specified by coordinates ($i,~j$), where $i$ is the position along the x-axis of length $L_{x}$ vertices and $j$ is the position along the y-axis of width $L_{y}$ vertices. The total number of vertices is $N_{T}=L_{x} L_{y}$. In Figure~\ref{ab40}a, strips of spin variables $s_{i}^{j}$ in the y-axis are referred to as $L_y$-mer's, where $s_{i}^{j} = -1$ or $+1$ for Ising-type models and $0$, $1$, or $2$ for Potts models. The critical nucleus can be represented by a column of $L_y$ proteins on the strip lattice. The proteins in the nucleus could also participate in inter-protein interactions other than the polymerizing interaction $K$ in the y-direction. These interactions can be described by using the sheet and helix interactions from the filament model, plus the free energy introduced above, $B>0$, that quantifies the inter-filament interactions between two sheet proteins. 

%%%%%%%%%%%%%%%%%%%%%%%%%%%%%%%%%%%%%%%%
	\begin{figure*}
	\centering
	\includegraphics[width = 350 pt  ]{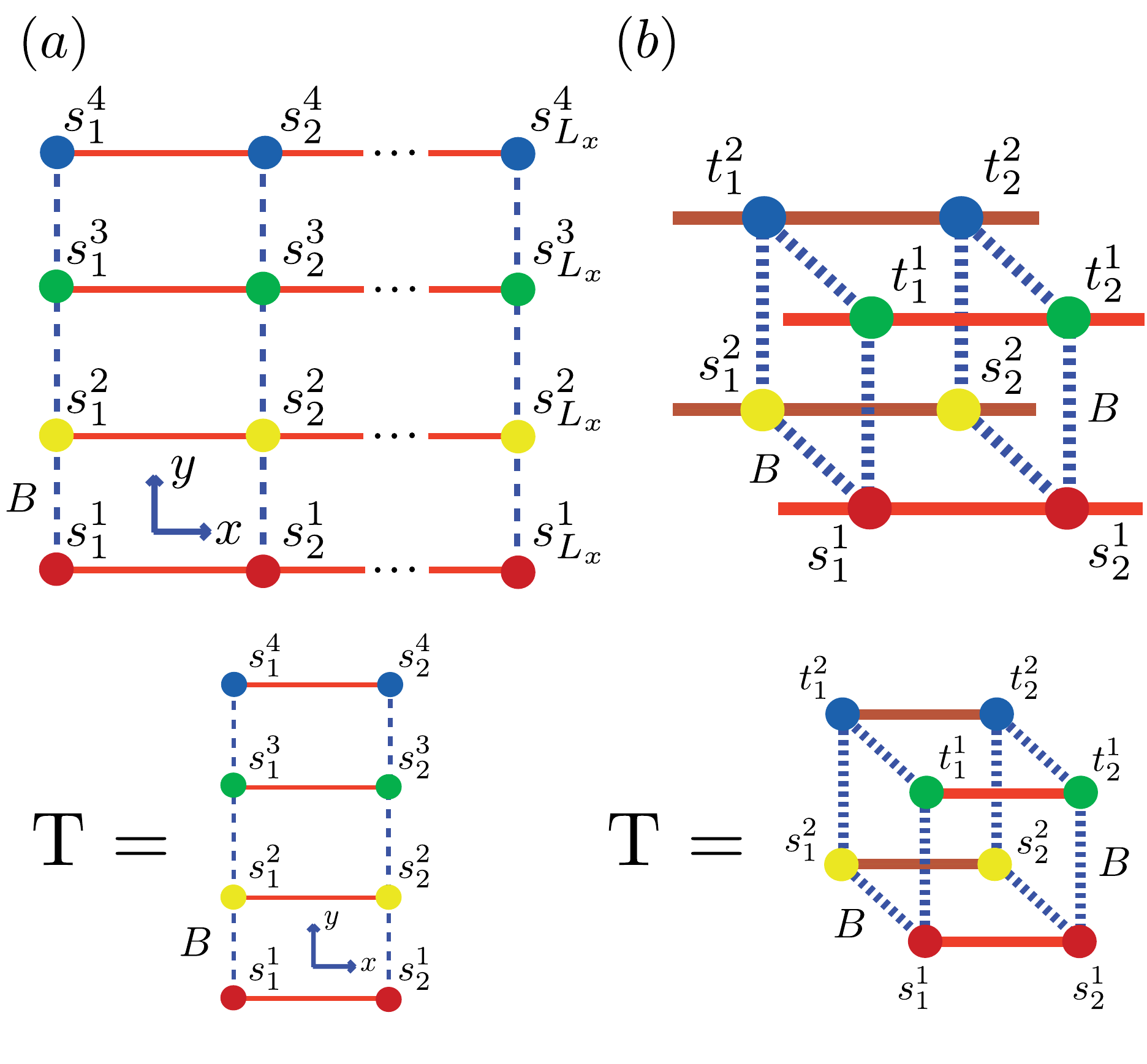}
	\caption{({\bf a}) Lattice representation and transfer matrix of $\alpha$-synuclein fibrils composed of 4 filaments; ({\bf b}) The fibril for $A\beta$ is composed of two protofibrils, each represented by a strip lattice and spin-variables $s$ and $t$. The strips are stacked in-register along the z-axis, and the parameter $B$ (dashed, green) lines represents the bonds between sheet proteins in protofibrils or fibrils. } %$\alpha$S figure adapted from Fink, Acc. Chem. Res. 2006, 39, 628-634.}
	\label{cube}
	\end{figure*}
%%%%%%%%%%%%%%%%%%%%%%%%%%%%%%%%%%%%%%%%

The interactions between the proteins in these aggregates is modeled similarly to the 2-helix chain model for proteins proposed by Skolnick~\cite{skolnick} and others~\cite{hausrath, qian, ghosh_dill, badasyan}, which use Zimm-Bragg or Lifson-Roig (LR) parameters to quantify the inter-chain interactions between residues in independent chains. When the inter-chain interactions between two helical residues are made zero, for example, the partition function reduces to a direct product of Zimm-Bragg~\cite{skolnick} (or LR~\cite{hausrath, qian}) transfer matrices. Since the model for aggregates proposed here uses a strip of a finite 2D lattice, as illustrated in Figure~\ref{cube}a, the two-protein case studied by Skolnick can be considered as a special case when considering protein folding instead of aggregation. The lessons learned from the folding models can guide us in constructing aggregation models.

%%%%%%%%%%%%%%%%%%%%%%%%%%%%%%%%%%%%%%%%
	\begin{figure}%f6
	\centering
	\includegraphics[width = \columnwidth  ]{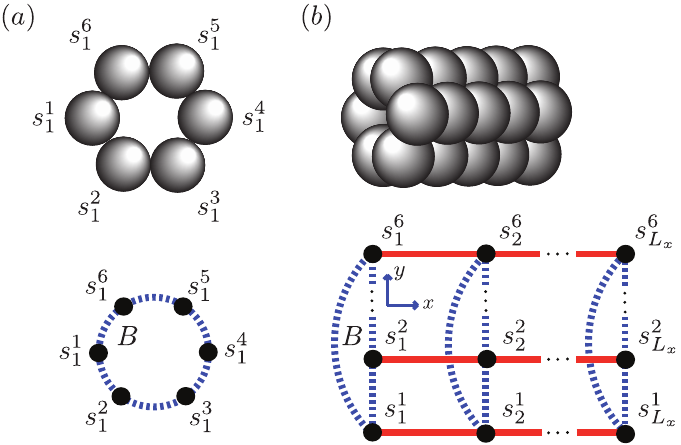}
	\caption{({\bf a}) A strip lattice with periodic boundary conditions in the {\em y}-axis is used to model the nucleus for A$\beta$(1-42) as a hexamer. The strip has $L_y$ monomers per column in general. $B$ (blue, dotted lines) is the free energy contribution from the interaction between proteins $s_{i}^{j}, s_{i}^{j+1}$ that are both sheet along the {\em y}-axis; ({\bf b}) Oligomers aggregate along the x-axis with a total of $L_x$ sites, where $L_x\rightarrow \infty$ is the thermodynamic limit. }%R (red, solid lines) is the nearest-neighbor interaction between proteins $s_{i}^{j}, s_{i+1}^{j}$ along the x-axis.  }
	\label{ab42fig}
	\end{figure}
%%%%%%%%%%%%%%%%%%%%%%%%%%%%%%%%%%%%%%%%

As mentioned, the nucleus is the smallest equilibrium aggregate in our formulation. The next smallest aggregate occupies the first two columns of the strip lattice, and contain 2$L_y$ number of proteins, then $4 L_y$ number of proteins, and so on. The interactions between $L_y$-mers along the x-axis can be described by generalizing the effective Hamiltonian for the 1D aggregation model, that is, $P_1>0$ and $P_2>0$ represent the interaction between helical and sheet proteins, respectively, and their nearest-neighbors in the aggregate. The total effective Hamiltonian for aggregates on a strip lattice with boundary conditions can be written as: 
\begin{equation}
\label{h_proto2}
\begin{split}
-\beta H_{2D} &= \sum_{j=1}^{L_y} H_{fil}(j) + B \sum_{i=1}^{L_x-1} \sum_{j=1}^{L_y-1} \delta(s_{i}^{j}, +2) \delta(s_{i}^{j+1}, +2) \\ & \quad + (L_y-1) L_x K \end{split}
\end{equation} 
\normalsize
where $H_{fil}$ is given by Equation~(\ref{hamil_hcs}) upon substituting $r_i \rightarrow s_i^j$ and where $s_{i}^{j}$ can assume the values of $0,~1,~2$ for coil, helix, or sheet monomers, respectively. Additionally, $B$ is the lateral binding interaction between two proteins from different filaments. The third term involving $K$ are the polymerizing interactions between proteins in both the {\em x} and {\em y} directions on the quasi-1D lattice. We assumed any polymerizing interactions between proteins on a quasi-1D lattice were equal in magnitude in order to keep the number of parameters used in the model to a minumum. It is similar to the parameter $K$ in Equation~(\ref{dutch_energy}), which took into account the polymerizing interaction between two adjacent proteins on the 1D lattice. For the case $q=2$ and $L_y=2$ of Equation~(\ref{h_proto2}), the Hamiltonian for sheet-coil protofibrils can be explicitly written as: 
\begin{equation}
\begin{split}
-\beta H_{2D}(s) &= H_{fil}(s^1) + H_{fil}(s^2) \\ & \quad + B \sum_{i=1}^{L_x-1} \delta(s_{i}^{1}, +2) \delta(s_{i}^{2}, +2) + L_x K \end{split}
\end{equation}
where $2D$ refers to aggregates on a strip lattice. The transfer matrix can then be written as:
\begin{equation}
\label{transfer_proto} 
T_{2D} = \left(
\begin{array}{cccc}
 1 & \sqrt{\sigma _1} & \sqrt{\sigma _1} & \sigma _1 \\
 s_1 \sqrt{\sigma _1} & s_1 & s_1 \sigma _1 & s_1 \sqrt{\sigma _1} \\
 s_1 \sqrt{\sigma _1} & s_1 \sigma _1 & s_1 & s_1 \sqrt{\sigma _1} \\
 s_1^2 \sigma _1 & s_1^2 \sqrt{\sigma _1} & s_1^2 \sqrt{\sigma _1} & s_1^2b
\end{array}
\right)
\end{equation}
where $b\equiv \exp(B)$. As mentioned, in the limit when the inter-filament interactions between two sheet proteins $B$ $\rightarrow0$, the transfer matrix given by Equation~(\ref{transfer_proto}) decomposes into a direct product of transfer matrices given by Equation~(\ref{ising}) for 1D filaments of length $N$ along the {\em x}-axis~\cite{Dutch2006}. Moreover, the limit in which the sheet-coil interfacial interactions $R$ $\rightarrow$ 0 (or $\sigma_1 \rightarrow 1$) yields independent strips parallel to the {\em y}-axis. The $2^{L_y} \times 2^{L_y}$ transfer matrix is also symmetric with respect to the sheet-coil interfacial interaction $R$. This fact is analogous to the 1D case, where the transfer matrix was symmetric in $\sigma_1$, $\sigma_2$ and $\sigma_3$. The partition function for aggregates on the strip lattice can be calculated by plugging the eigenvalues, $\lambda_{2D, i}$, of Equation~(\ref{transfer_proto}) into Equation~(\ref{partition_function})  and specifying boundary conditions. The result is:  
\begin{eqnarray}
\label{z2d}
Z_{2D} &=& k^{L_y(2L_x-1)-L_x} \sum_{i=1}^{N_{\lambda_{2D}}} c_{i} \lambda_{2D, i}^{L_x-2}.
\end{eqnarray}
where, again, $c_i$ are determined by boundary conditions and $k$ was defined in Equation~(\ref{tftftf}).

Using the ZB formalism, we can also model fibrils in the {\em x}-, {\em y}- and {\em z}-directions using a quasi-1D lattice in 3D. For example, two or more filaments could join to form a protofibril or fibril, as illustrated in Figure~\ref{cube} for two simple geometric configurations. For simplicity, we study the case where two 2D aggregates represented by strip lattices, as depicted in Figure~\ref{cube}b, are the same geometrical shape and are stacked one on top the other, in-register. The spin variable associated with one of the aggregates is denoted by $s$, while the spin variables associated to the second aggregate is denoted by $t$. The effective Hamiltonian for the fibril model in Figure~\ref{cube}b, referred to as the ``cube'' model, may be written using  the strip model for aggregates as: 
\begin{small}
\begin{eqnarray}
-\beta H_{3D}&=& -\beta H_{2D}(s) -\beta H_{2D}(t)  \\ 
&+& B \sum_{i=1}^{N}  \delta( s_{i}^{1}, +2 ) \delta( t_i^1, +2) \delta( s_{i}^{2}, +2 ) \delta( t_{i}^{2},+2 ) + L_x L_z K \nonumber
\end{eqnarray}
\end{small}
where we assume that the interaction between any two sheet proteins from adjacent aggregates is also quantified by the free energy $B>0$. Additionally, the aggregates can now polymerize in any direction. To help keep the number of parameters used in the model to a minimum, we assume the polymerizing interactions between two adjacent proteins in the z-direction has the same strength as the polymerizing interactions, $K$, in the x and y directions. The corresponding transfer matrices for each model for fibrils are found just as they were for the 1D and strip models discussed earlier. The result is: 
\begin{eqnarray}
\label{z3d}
Z_{3D} &=& k^{-L_y+L_x \left(-1+2 L_y+L_z\right)} \sum_{i=1}^{N_{\lambda_{3D}}} c_{i} \lambda_{3D, i}^{L_x-2}
\end{eqnarray}

In general, the transfer matrix for the 2D model has dimension $q^{L_y} \times q^{L_y}$, while the 3D model has dimension $q^{2 L_y} \times q^{2 L_y}$, where we assumed that the two protofibrils composing the fibril contain $L_y$ number of filaments. The transfer matrices for both the 2D and 3D models are illustrated in Figure~\ref{cube}a and Figure~\ref{cube}b, respectively. Since the Potts model is used, $q$ is 2 for sheet-coil, helix-coil, or helix-sheet models, and 3 for the helix-sheet-coil model. 
%
%%%%%%%%%%%%%%%%%%%%%%%%%%%%%%%%%%%%%%%%

\subsection{Dilute Thermodynamic Averages}

To compare with experiments, some average properties of the dilute system of monomers and aggregates can be defined. The total number density for the strip or cube model for fibrils can be written~as: 
\begin{equation}
\rho(L) = Z(L) \exp\left( \mu L \right)
\end{equation}
where $Z$ can be given by Equation~(\ref{z2d}) or (\ref{z3d}) for 2D or 3D aggregates, respectively, and $L$ is the total number of proteins in aggregates. In the 2D case, $L = L_x L_y$. The average fraction that an aggregate is helix ($i=1$) or sheet ($i=2$) can be defined as:
\begin{equation}
\label{thetaks}
\langle \theta_i \rangle \equiv \frac{ \displaystyle \sum_{L_x=1}^{\infty} \theta_i (L) L \rho(L) }{ \displaystyle \sum_{L_x=1}^{\infty} L \rho(L) }
\end{equation}
where the {\em x}-axis is the axis of propagation of the aggregate, and $\theta_i$ can be calculated for any of the aggregate species discussed earlier by computing Equation~(\ref{general_thermo}). Equation~(\ref{thetaks}) can be used as a fit function for CD spectral data points. The average degree of polymerization of an aggregate growing in the {\em x}-direction can also be defined as:
\vspace{6 pt}
\begin{equation}
\langle L_x \rangle \equiv \frac{ \displaystyle \sum_{L_x=1}^{\infty} L \rho(L) }{ \displaystyle \sum_{L_x=1}^{\infty} \rho(L) }
\vspace{6 pt}
\end{equation}
and is directly related to the length of the fibrils. The expressions for $\langle \theta_1 \rangle$ and $\langle L \rangle$ can be obtained for systems of $\alpha$-synuclein ($\alpha$S) and A$\beta$(1-40) aggregates. The $\alpha$-synuclein fibril is modeled by placing the proteins in the aggregates onto the $L_{y=4}$ strip lattice, as illustrated in Figure~\ref{cube}a. Thus, the average length of these fibrils is then: 
\begin{equation}
\label{aveleng}
\langle L \rangle = \frac{ \langle L_x \rangle }{4} 
\end{equation}
which can be used to fit the AFM measurements of the average lengths of the fibrils. This relation also holds for the average length of a fibril described by the cube lattice model as depicted in Figure~\ref{cube}b for the A$\beta$(1-40) fibrils. As an example, we calculate these quantities explicitly for a system of equilibrium A$\beta$(1-40) aggregates using a sheet-coil model. Specifically, for a system where 1D filaments, $L_y=2$ strip aggregates, or 3D cube aggregates could be present at equilibrium, we first define:
\begin{eqnarray}
\rho_{A\beta} &\equiv& \sum_{L_x=1}^{\infty} \rho(L_x) + \rho(2 L_x) + \rho(4 L_x) \nonumber \\ 
&=& z + k z^2 \langle i | f \rangle_{1D}  + \sum_{i=1}^{N_{\lambda_{1D}}} \frac{ k^{2} z^{3} x_{i} \lambda_{i}}{1-k z \lambda_{i}}  \nonumber \\
&+& z^2 + k^4 z^4 \langle i | f \rangle_{2D} + \sum_{j=1}^{N_{\lambda_{2D}}} \frac{ k^{7} z^{6} x_{j} \lambda_{j}}{1-k^3 z^2 \lambda_{j}} \nonumber \\
&+& z^4 + k^{12} z^8 \langle i | f \rangle_{3D}  + \sum_{l=1}^{N_{\lambda_{3D}}} \frac{ k^{13} z^{12} x_{l} \lambda_{l}}{1-k^5 z^4 \lambda_{l}} 
\end{eqnarray}
where the fugacity $z\equiv \exp(\beta \mu)$ and in each sum $\lambda_k$ is the $k$th eigenvalue of the 1D, 2D, or 3D transfer matrix for filament, strip, or cube models, respectively. The A$\beta$(1-40) aggregates considered here at equilibrium are illustrated in Figure~\ref{fitpathways}. Additionally, $x_k$ is the $k$th term of the expression $\langle i | f \rangle$, where $| i \rangle$ and $| f \rangle$ are the specified boundary conditions. The sums computed converge only if $k z \lambda_{j} < 1$ for all $j$. Details on boundary conditions for ZB-type models can be found in \cite{dutch_2003}. By using Equation~(\ref{con_mass}), $\phi=m_{tot}/V$ can also be written explicitly for A$\beta$(1-40) aggregates as: 
\vspace{-3 pt}
\begin{eqnarray}
\phi &\equiv& \sum_{L_x=1}^{\infty} L_x \rho(L_x) + 2 L_x \rho(2 L_x) + 4 L_x \rho(4 L_x)  \\
&=& z + 2 k z^2 \langle i | f \rangle_{1D} + \sum_{j=1}^{N_{\lambda_{1D}}} \frac{ (k^{2} z^{3} x_{j} \lambda_{j})(3-2 k z \lambda_{j})}{(1-k z \lambda_{j})^2} \nonumber \\
&+& 2 z^2 + 4 k^4 z^4 \langle i | f \rangle_{2D} + \sum_{j=1}^{N_{\lambda_{2D}}} \frac{ 2 (k^{7} z^{6} x_{j} \lambda_{j})(3-2 k^3 z^2 \lambda_{j})}{(1-k^3 z^2 \lambda_{j})^2} \nonumber \\
&+& 4 z^4 + 8 k^{12} z^8 \langle i | f \rangle_{3D} + \sum_{j=1}^{N_{\lambda_{3D}}} \frac{ 4 (k^{13} z^{12} x_{j} \lambda_{j})(3-2 k^5 z^4 \lambda_{j})}{(1-k^5 z^4 \lambda_{j})^2} \nonumber
\end{eqnarray}

Now the average lengths of the aggregates can be computed. Next, the average fraction of the aggregates that are sheet, $\langle \theta_2 \rangle$, is calculated for the A$\beta$(1-40) model. By plugging Equation~(\ref{theta}) for filament, strip, and cube aggregates into Equation~(\ref{thetaks}), $\langle \theta_2 \rangle$ can be written as: 
\begin{widetext}
\begin{eqnarray}
\label{thetacube}
\langle \theta_{2} \rangle_{A\beta} &=& \frac{s_2}{\phi} \sum_{k=1}^{N_{1D}} \frac{ x_k \frac{\partial \lambda_k}{\partial s_2} k^2 z^3 \left( 3 - 2 k z \lambda_k \right)}{ \left(1-k z \lambda_k \right)^2 } + \frac{ \frac{\partial x_k}{\partial s_2}  k z^2 \left(k z \lambda_k + 2(-1+k z \lambda_k) \log(1-k z \lambda_k) \right) }{ 1-k z \lambda_k}  \nonumber \\
&+& \frac{s_2}{\phi} \sum_{k=1}^{N_{2D}} \frac{ 2 x_k \frac{\partial \lambda_k}{\partial s_2} k^7 z^6 \left( 3 - 2 k^3 z^2 \lambda_k \right) }{ (1-k^3 z^2 \lambda_k)^2} + \frac{ 2 \frac{\partial x_k}{\partial s_2} k^7 z^6 \lambda_k }{ 1-k^3 z^2 \lambda_k} - 4 \frac{\partial x_k}{\partial s_2} k^4 z^4 \log( 1 - k^3 z^2 \lambda_k ) \nonumber \\
&+& \frac{s_2}{\phi} \sum_{k=1}^{N_{3D}} \frac{ 4 x_k \frac{\partial \lambda_k}{\partial s_2} k^{13} z^{12} \left( 3 - 2 k^5 z^4 \lambda_k \right) }{ (1-k^5 z^4 \lambda_k)^2} \nonumber \\
&+& 4 \frac{\partial x_k}{\partial s_2} k^8 z^8 \lambda_k \left(-1 + \frac{1}{1-k^5 z^4 \lambda_k} - 2 \log( 1-k^5 z^4 \lambda_k ) \right) 
\end{eqnarray}
\end{widetext}
where $s_2$ is the sheet propagation parameter. The procedure for finding $\langle \theta_2 \rangle$ is quite general, and works for all of the transfer matrices that we have considered in this model. 

%%%%%%%%%%%%%%%%%%%%%%%%%%%%%%%%%%%%%%%%
	\begin{figure*}%f8
	\begin{center}
	\includegraphics[width = 350 pt  ]{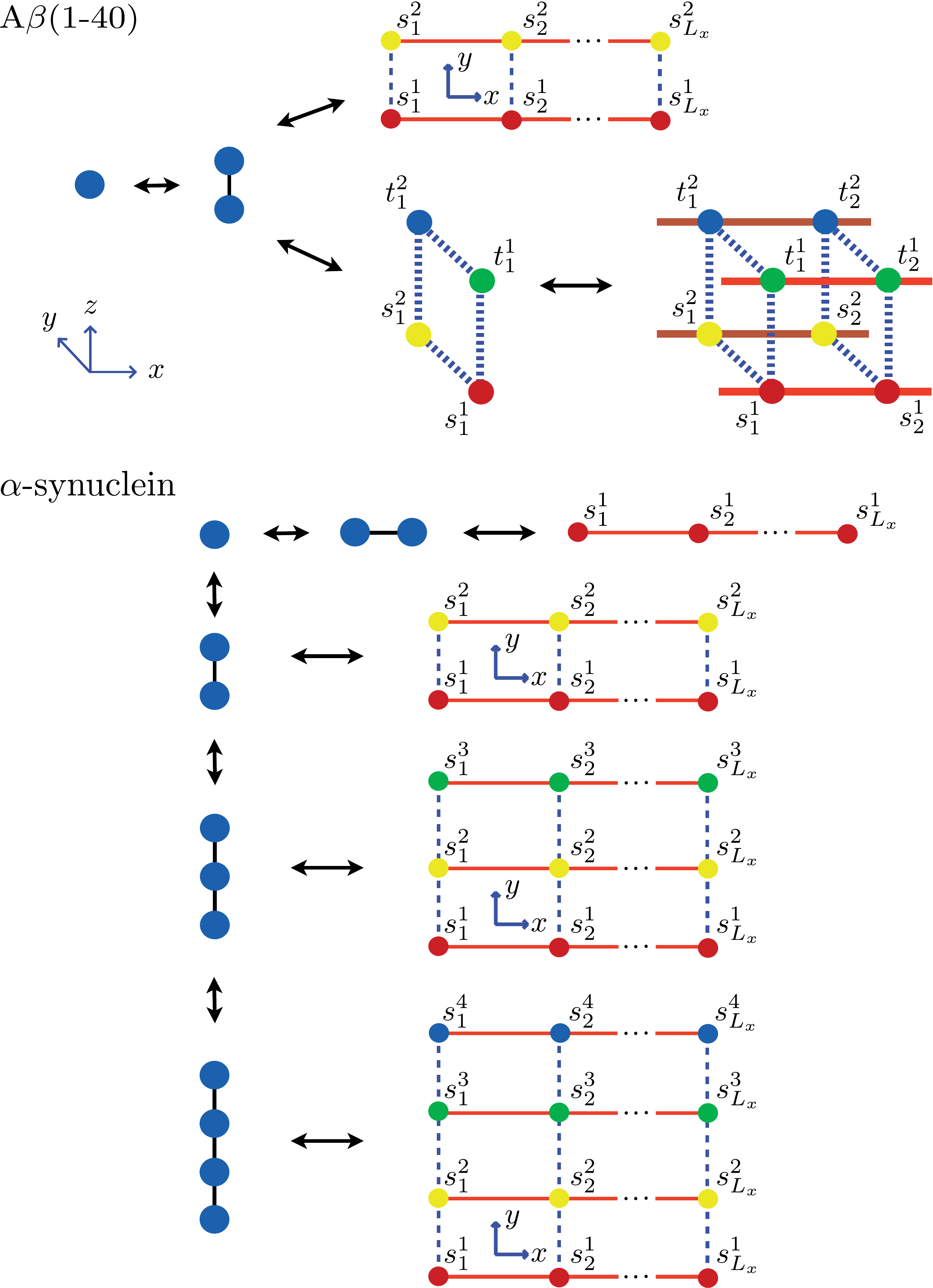}
	\caption{Partial lists of chemical species that may exist in dynamic equilibrium with fibrils for A$\beta$(1-40) ({\bf top}) and $\alpha$-synuclein ({\bf bottom}). In the A$\beta$ model, the different types of aggregates that could be present at equilibrium are 1D filaments, strips of length $L_y=2$ that represent protofibrils, and 3D cubes that represent fibrils. The cubes are composed of two identical proto-fibrils stacked in-register. In the model for $\alpha$-synuclein aggregates, \mbox{$L_y$ = 1, 2, 3,} and 4 strip lattices are used to describe the aggregates at equilibrium, with the $L_y$ = 4 strip lattice representing the fibril. For both A$\beta$(1-40) and $\alpha$-synuclein, we assumed~$n_c =~2$. }
	\label{fitpathways}
	\end{center}
	\end{figure*}
%%%%%%%%%%%%%%%%%%%%%%%%%%%%%%%%%%%%%%%%

\subsection{Comparison to Experiment} 

In this subsection, our ZB-like model predictions are compared to the experimental results for the CD spectra of A$\beta$(1-40) fibrils~\cite{terzi} and the AFM measurements of the lengths of $\alpha$-synuclein fibrils~\cite{dutch_2008}. The CD and the AFM measurements were made at various initial mass concentrations of each protein, when the fibrils had reached a steady state. For the fit of the CD data, we used as our fit function the total fractional amount of sheet structure in all of the aggregates, $\langle \theta_{2} \rangle_{A\beta}$, given by Equation~(\ref{thetacube}).  The proteins at the boundaries of aggregates could be coil or sheet for 1D, 2D, and 3D lattices. The fit and the values for $P_2$, the sheet interaction free energy, $K$, the free energy describing the polymerization of the aggregate in any direction, $R_2$, the sheet-coil interfacial free energy, and  $B$, and lateral binding free energy for the $q=2$ sheet-coil model are given in Figure~\ref{data_fits_cdafm}a. The fit parameters are then used to predict the average length of the fibrils in the system, $\langle L \rangle$, which illustrated in Figure~\ref{data_fits_cdafm}b. 

%%%%%%%%%%%%%%%%%%%%%%%%%%%%%%%%%%%%%%%%
	\begin{figure*}
	\centering
	\includegraphics[width = 420 pt  ]{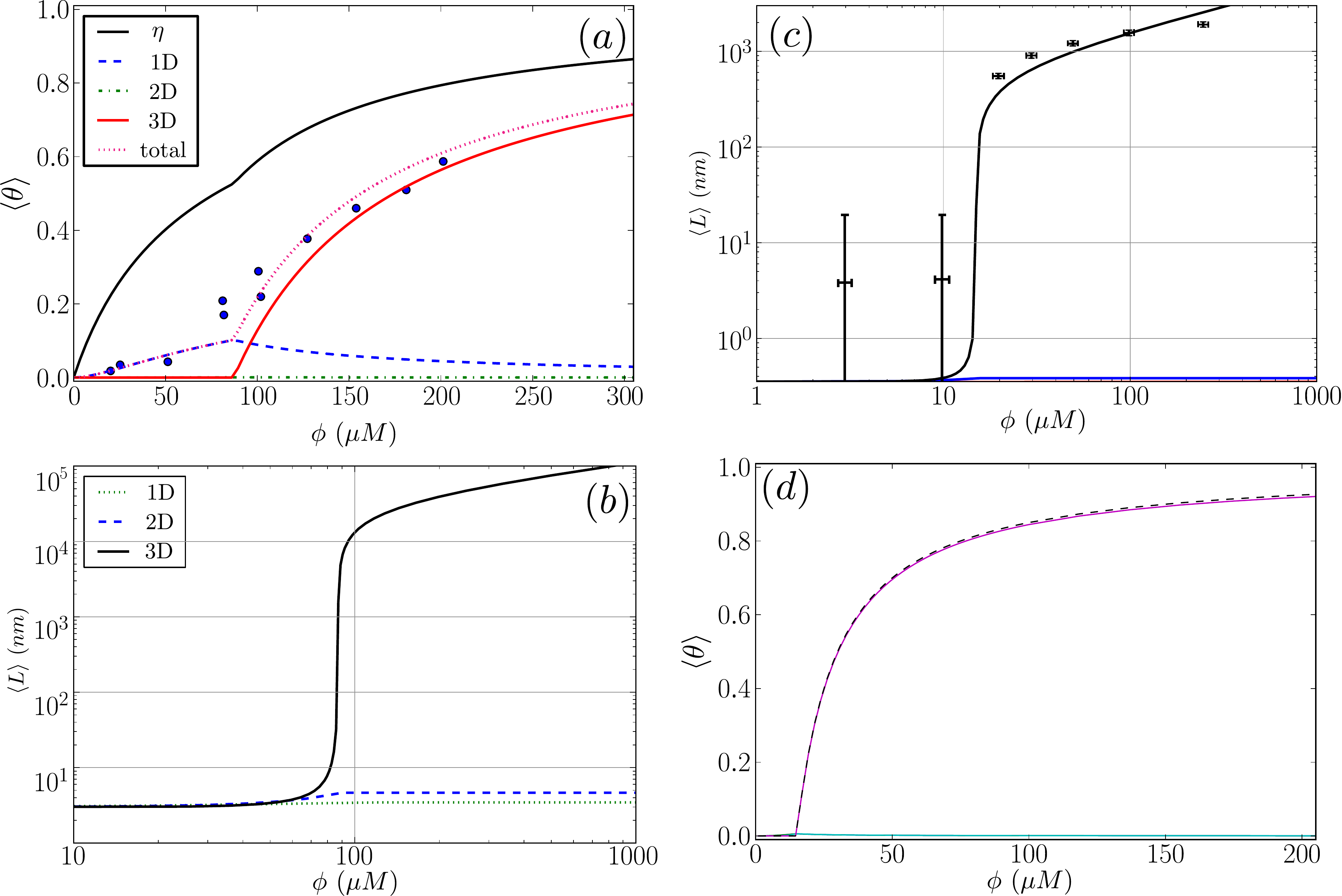}
	\caption{({\bf a}) Plot of $\langle \theta_2 \rangle$ for 1D, 2D, and 3D structures in the A$\beta$(1-40) model. Black dots represent the CD data from Terzi, {\em et al.}~\cite{terzi}, where we the total fraction of sheet proteins in aggregates of any species; ({\bf b}) Predicted average lengths, $\langle L \rangle$, of the A$\beta$(1-40) fibrils using the fit parameters found in ({\bf a}); In plot ({\bf c}), the AFM data for the $\alpha$-synuclein fibrils is plotted as black dots, along with the fit function $\langle L \rangle$~\cite{dutch_2008}. We fit $\langle L \rangle$ using the $L_y = 4$ strip lattice model; In ({\bf d}), $\langle \theta_2 \rangle$ for $\alpha$-synuclein (solid, purple curve) is compared with $\rho_{fib} / \phi$ (dashed, black curve) by using the fit parameters found in ({\bf c}); In ({\bf a}) and ({\bf b}), the fit parameters for the A$\beta$(1-40) model were: $P_1 = 7.41RT$, $B = 1.4RT$, $R_2 = -2.47RT$, and $K = 0.45RT$. \protect In ({\bf a}), $\eta = 1 - (z +z^2 + z^4)/\phi$; In ({\bf c}) and ({\bf d}), the fit parameters for the $\alpha$-synuclein model were $P_1 = K = 2.7 RT$, $B = 1.95RT$ and $R_2 = -1.64 RT$. }
	\label{data_fits_cdafm}
	\end{figure*}
%%%%%%%%%%%%%%%%%%%%%%%%%%%%%%%%%%%%%%%%

For the $\alpha$-synuclein model, where $L_y$ = 1, 2, 3 and 4 strip aggregates could be~present, we fit the $L_y = 4$ contribution from Equation~(\ref{aveleng}) to the AFM average length data for the fibrils.~The $\alpha$-synuclein aggregates at equilibrium are illustrated in Figure~\ref{fitpathways}. The fit is illustrated in Figure~\ref{data_fits_cdafm}c. The fit parameters are then used to predict the~total fraction of aggregates that are sheet, which can be written for the $\alpha$-synuclein  model as: 
\begin{equation}
\label{thetasyn}
\langle \theta_{2} \rangle_{\alpha S} = \langle \theta_{2} \rangle_{ L_y=1} + \langle \theta_{2} \rangle_{L_y=2} + \langle \theta_{2} \rangle_{L_y=3} + \langle \theta_{2} \rangle_{L_y=4}
\end{equation}
where each contribution in Equation~(\ref{thetasyn}) can be calculated using transfer matrices derived from Equation~(\ref{h_proto2}) for $L_y=1,...,4$. Equation~(\ref{thetasyn}) is illustrated in Figure~\ref{data_fits_cdafm}d. The model predicts the average length data pretty well (Notice the log scales used in the plot. At the low coverage, the predicted points fall actually within the error bars), as we should have expected because the other variations of the Ising-ZB model fit the data well~\cite{Dutch2006, dutch_2008}. The resulting predictions for $\langle \theta_{2} \rangle_{\alpha S}$ illustrate that the concentration at which the fibril concentration takes off is around 15 $\upmu$M, again, as we should  have expected~\cite{dutch_2008}.

The fit of $\langle \theta_{2} \rangle_{A\beta}$ to the CD data predicts that the fibrils are held together tightly due to the relatively large value of the sheet-interaction between proteins in the aggregates, $P_1$, and to a lesser extent on the binding between filaments as quantified by $B$. The fitted value for the sheet-coil interface free energy, $R_2$, indicates that the interfacial tension between sheet and coil regions in the aggregates is modest. However, the fibril concentrations do not really increase from zero until nearly 100 $\upmu$M according to the model predictions, but $\beta$-rich fibrils have been observed at lower concentrations as seen in Figure~\ref{data_fits_cdafm}a. The fit could be improved by considering other types of models for the A$\beta$ fibrils and as well as different boundary conditions. 

The model predictions for the strip model of $\alpha$-synuclein fibrils seem to agree with the experimentally determined average lengths as illustrated in Figure~\ref{data_fits_cdafm}c where the boundary conditions were set so that the ends of fibrils could be sheet or coil proteins. Additionally, as illustrated in Figure~\ref{data_fits_cdafm}d, the model for synuclein fibrils predicts that the sheet-coil transition of proteins in fibrils largely drives the polymerization process, where $\rho_{fib} / \phi$ and $\langle \theta_{2} \rangle_{L_y=4}$ give nearly the same result for the concentrations used in the AFM experiments. 

The fits of the AFM data done by van Raaij {\em et al.}~\cite{dutch_2008} and Schmit {\em et al.}~\cite{Schmit2011} needed only 2 parameters, compared to 3 in the present model. When compared with van Raaij's fit of the AFM data, our model predicts that the probability that fibrils contain sheet structure is high once overcoming a sheet-coil free energy barrier $R_2$, whereas in van Raaij's model prediction, the free energy barrier between adjacent sheet and coil proteins in the aggregates does not seem to be present. A finite contribution from $R_2$ means the fibrils will have longer stretches of sheet content when compared to cases when $R_2$ is closer to zero when there is little or no penalty between coil to sheet regions in the fibrils. Additionally, our model predicts that the inter-filament interactions, $B$, are slightly weaker than the interactions between sheets along the axis of growth of the fibril, whereas van Raaij's fit is the other way around: the inter-filament interactions are stronger than those between proteins along the fibril axis of growth. 

When fitting the A$\beta$(1-40) CD data, our model predicts that the value of the polymerizing interaction between proteins on a quasi-ID lattice, $K$, is small and could be due to modeling uncertainty, thus the number of parameters needed to fit the CD data could be less. The fact that not all adjustable parameters were required to fit CD and AFM data suggests that the model Hamiltonians introduced throughout the paper may be simplified to describe only the relevant interactions quantified by the non-zero fit parameters. It may also imply that the fibrils are mainly held together by a few types of energetic interactions, for example, the inter-filament interaction $B$ for A$\beta$(1-40) had a finite contribution in the Hamiltonian. Other interactions described by the Hamiltonians could be non-existent or very small. More detailed experimental results are needed to discern the correctness of the models in predicting the values of interaction energy parameters, which in turn describe the dominant interactions within the~fibrils. 

As mentioned, the data sets were fit using various boundary conditions, and the open boundary case (proteins could adopt either conformation at the boundaries) was found to be the best choice for fitting for the average lengths of the $\alpha$-synuclein fibrils. The CD fits could not be shown to be dependent on certain boundary conditions since there is currently no AFM measurements of the fibrils to compliment the CD data. This means we could fit the CD data using most choices for boundary conditions, including the case where all proteins at the ends of fibrils are in the coil conformation. However, for some choices of boundary conditions~\cite{dutch_2003, Dutch2006}, the corresponding average length predictions yielded unreasonable lengths (not shown) for protein aggregates {\it in vivo}. 

\section{Grand Canonical Approach}

The models for fibrils discussed so far do not take into account interactions between protein and solvent, or some free energy that would be associated with nucleus formation. The ZB model for aggregation can be extended to take into account these phenomena by using a grand-canonical model. We summarize several main differences between canonical and grand-canonical approaches: (1) in the grand canonical model, aggregates of all sizes are included; (2) an aggregate phase and solution phase are in equilibrium; (3) chemical potentials can be used relating the solution phase as well as the aggregate~phase~\cite{Schreck2011}. 

The solution phase is defined by specifying the chemical potential for protein monomers in the solution can be written~\cite{Ferrone2006, Ferrone1997, Hill1987} as:
\begin{equation}
\label{chemsoln}
\mu_{soln} = \mu_{ST} + \mu_{SR} + R T \ln c
\end{equation}
where the subscript ``{\em S}'' stands for solution, $\mu_{ST}$ and $\mu_{SR}$ are the free energy contributions arising from the translational and rotational motion of monomers moving in solution, respectively, and  $c$ is the concentration of monomers in solution. The aggregate phase is defined by specifying the chemical potential of the aggregates, $\mu_{agg}$, by assuming a crystalline approximation so that $\mu_{agg}$ can be written~as~\cite{Abraham1974}:
\begin{equation}
\label{chemagg}
\mu_{agg}  = \mu_{PC} + \mu_{PV}
\end{equation}
where ``{\em P}'' stands for polymers of proteins. $\mu_{PC}$ is the free energy contribution arising from the contact interactions between proteins in aggregates, which may vary for different monomer organizations in the aggregates. We assume this term also includes the conformational entropy of the backbone and side-chains. The term $\mu_{PV}$ is the free energy arising from the proteins vibrating about their equilibrium positions, but not molecular internal vibrations within the proteins~\cite{Abraham1974, Ferrone2006}. When the phases are at equilibrium, the chemical potentials for each phase are equal: 
\vspace{-6 pt}
\begin{equation}
\label{equib}
\mu_{agg} = \mu_{soln}
\end{equation}

With the simple statistical mechanical model summarized in the sections to follow, we can relate the chemical potential contribution from the protein interactions in aggregates, $\mu_{PC}$, to the experimental concentration of protein in solution via Equation~(\ref{equib}).

As a first step in generalizing the canonical effective-Hamiltonian models presented earlier, the  free energy $A$ is introduced to quantify the entropic penalty needed to nucleate the aggregate, {\em i.e.}, the first column of a strip lattice that contains protein aggregates. This free energy may also be viewed as a boundary between proteins and solvent. The aggregate phase then assumes strip or cube lattices may be occupied by aggregates and any other species, including solvent clusters. 

To write down an effective Hamiltonian that can include the free energy $A$, the 1D or quasi-1D lattices used in constructing fibrils can be generalized to allow solvent clusters to occupy the lattice sites. For example, in Figure~{\ref{weight}, a square represents a solvent cluster, whereas circle represents protein. Both solvent and proteins can occupy sites along 1D or quasi-1D lattices. By introducing a lattice gas model into the aggregate phase, a Potts Hamiltonian for the 1D lattice that quantifies the interactions between helix, sheet, or coil proteins and solvent can be written as~\cite{Schreck2011}:
{\small
\begin{eqnarray}
\label{latticegas}
-\beta \mathcal{H}_{fil} &=& -\beta \mathcal{H}_{pp} - \beta \mathcal{H}_{ps}^{n_c} \\
\label{pro_pro}
-\beta \mathcal{H}_{pp} &=& \sum_{i=1}^{N_{T}-1} \left\{ P_1~\delta( t_{i}, 1 ) + K - R_1\chi(t_{i},t_{i+1}) \right\} n_{i} n_{i+1} \nonumber \\
&-& \sum_{i=1}^{N_{T}-1} R_1\chi(n_{i}, n_{i+1}) \left[ \delta(t_{i},1) n_{i} + \delta( t_{i+1}, 1) n_{i+1} \right] \\
-\beta \mathcal{H}_{ps}^{n_{c}} &=& - \sum_{i=1}^{N_{T}-n_{c}-1} A \chi( n_i, n_{i+n_{c}} ) \prod_{j=i+1}^{i+n_{c}-1} \delta( n_j, 1 ) 
\label{nucleus}
\end{eqnarray}}
where the lattice-gas variable $n_i=1$ refers to a protein occupied lattice site, and $n_i=0$ a solvent occupied site. Additionally, ``pp'' in $-\beta \mathcal{H}_{pp}$ refers to ``protein-protein'' interactions and ``ps'' in $-\beta \mathcal{H}_{ps}^{n_{c}}$ refers to ``protein-solvent'' interactions. The term $\chi( n_i, n_{i+n_{c}} )=1-\delta(n_i, n_{i+n_{c}})$ ensuring that there is solvent at site $i$ and a protein at $i+n_{c}$, or vice-versa.

\begin{figure}%f10
	\centering
	\includegraphics[width = \columnwidth ]{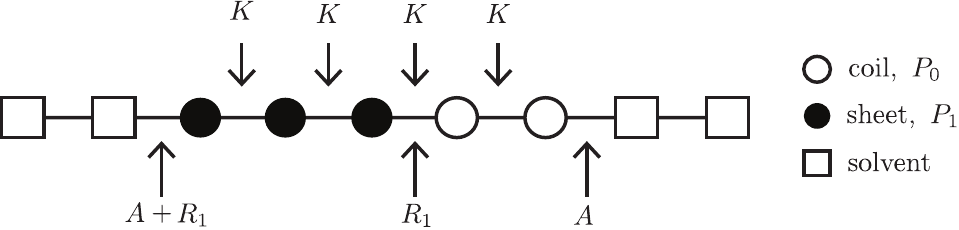}
	\caption{ Summary of protein conformation energies. A site could be occupied with a solvent cluster, denoted by $n=0$ (square), or a protein, $n=1$ (circles). Proteins may assume a particular conformation (sheet, black/solid circle; coil, white circle). A dilute $q=2$ Potts model for sheet-coil conformations is shown, where $n_{c}=1$ and the free energies $P_1$, $K$, $R$, and $A$ are illustrated.}
	\label{weight}
\end{figure}

Since the number of proteins on the lattice can fluctuate, this description of protein aggregation is described by using the grand canonical ensemble. The lattice-gas formalism, {\em i.e.}, Equation~(\ref{latticegas}), is able to describe a variety of elongation mechanisms including merging and fracturing of aggregates of different sizes along the 1D lattice. The partition function can be written as: 
\begin{equation}
\label{grand}
\mathcal{Q} = \sum_{\{t\}, \{n\}} \exp\left(-\beta \mathcal{H}_{fil} + \beta \mu_{PC} N_{p}\right) 
\end{equation} 
where $\beta \mu_{PC}$ is the dimensionless chemical potential arising from the contact and interfacial interactions between proteins in aggregates, and where the sum is performed over both spin and lattice-gas variables. Just like in the canonical models, $\mathcal{Q}$ may be solved for exactly by a transfer matrix $T$. A simple example illustrating $T$ for the case $n_{c}=1$ in a sheet-coil ($t_i=-1$ for coil, $t_i=1$ for sheet) system can be written~as: 
\begin{eqnarray}
\mathcal{T} &=&
\begin{tabular}{ c c  | c c c}
	& $t_{i+1}$ &  & $-1$ & $1$ \\
	& $n_{i+1}$ & $0$ & $1$ & $1$ \\
	$t_{i}$ & $n_{i}$ \  &  &  &  \\
  	\hline\noalign{\smallskip}
   	 & $0$~~ & $1$ & $\sqrt{\alpha}$ & $\sqrt{\alpha \sigma_1}$ \\
  	$-1$ & $1$~~ & $z\sqrt{\alpha}$ & $kz$  & $kz\sqrt{\sigma_1}$ \\ 
  	~$1$ & $1$~~ & $z\sqrt{\alpha \sigma_1}$ & $kzs_{1}\sqrt{\sigma_1}$ & $kzs_{1}$  \\
\end{tabular}
\end{eqnarray}
where $s_1$, $\sigma_1$, and $k$ were defined earlier and $\alpha\equiv\exp(-2A)$ is a new Zimm-Bragg-like parameter. Additionally, the fugacity is now defined as $z \equiv \exp(\beta \mu_{PC})$. 

The inter-filament interactions between two 1D filaments are treated using the same methodology introduced in earlier sections. In general, the Hamiltonian for an $L_x \times L_y$ strip lattice that includes inter-filament interactions can be written using the 1D Hamiltonian, Equation~(\ref{latticegas}), by changing the spin and lattice-gas variables $t_{i} \rightarrow t_{i}^{j}$ and $n_{i} \rightarrow n_{i}^{j}$, respectively, as:
\begin{eqnarray}
\label{two_filament}
-\beta \mathcal{H}_{strip}^{A} &=& - \sum_{j=1}^{L_y} \beta \mathcal{H}_{fil}(j) \nonumber \\
&+& F \sum_{i=1}^{N} \sum_{j=1}^{L_y-1} \delta(t_{i}^{j}, 1) \delta(t_{i}^{j+1}, 1) n_{i}^{j}n_{i}^{j+1} 
\end{eqnarray}
where $\mathcal{H}_{fil}(j)$ refers to the $j$th filament. For A$\beta$(1-40) $L_{y}=2$, as illustrated in Figure~\ref{ab40}b,c. The parameter $F$ quantifies the interaction energy between two sheet-linked proteins from adjacent filaments, and plays the same role as the free energy $B$ in earlier models for fibrils in this article. In our treatment $F>0$, the proto-fibrils and fibrils are more stable than single filaments. 

Since nucleation cannot in reality occur in 1D, we consider a similar model for aggregates that positions the nucleus along the y-axis, as shown in Figure~\ref{cube}a and Figure~\ref{ab40dilute}a. From this point of view the orientations of proteins in the nucleus are perpendicular to the direction of propagation ({\it x}-axis) of the fibrils, and the nucleus is now a multi-layer, 1D aggregate. The nuclei may assemble into proto-fibrils that grow longer on the quasi-1D lattice. An effective Hamiltonian for protein aggregation, including the quasi-1D nucleus, can be written: 

\small
\begin{eqnarray}
\label{hamilb}
-\beta \mathcal{H}_{strip}^{B} &=& - \sum_{j=1}^{L_y} \beta \mathcal{H}_{pp}(j) - \sum_{j=1}^{L_y-1} \beta H_y(j) - \beta H_{nuc} \\ 
- \beta H_y(j) &=& \sum_{i=1}^{N_{T}} \left\{ F~\delta( t_{i}^{j}, 1 )+ K - R_1\chi(t_{i}^{j} ,t_{i}^{j+1})\right\} n_{i}^{j} n_{i}^{j+1} \nonumber \\
&-& \sum_{i=1}^{N_{T}} R_1\chi(n_{i}^{j} , n_{i}^{j+1}) \left[  \delta(t_{i}^{j} ,1) n_{i}^{j} + \delta( t_{i}^{j+1}, 1) n_{i}^{j+1} \right]  \\ 
- \beta H_{nuc} &=& - \sum_{i=1}^{N_T - 1} A \prod_{j=1}^{L_y} \chi(n_{i}^{j} , n_{i+1}^{j}) \prod_{j=1}^{L_y-1} \delta(n_i^j, n_{i}^{j+1})
\end{eqnarray}
\normalsize
where $-\beta H_{pp}(j)$ is given by Equation~(\ref{pro_pro}) after substituting $t_{i} \rightarrow t_{i}^{j}$ and $n_{i} \rightarrow n_{i}^{j}$. In the y-direction we write analogous interactions, $-\beta H_y$, similar to those in the {\em x}-direction. Also included in the y-direction is the nucleus term containing the parameter $A$, which has the same meaning of surface energy as before. The effective Hamiltonians given by Equations~(\ref{two_filament}) and (\ref{hamilb}) are the most general forms of fibrils that we have considered so far.

%%%%%%%%%%%%%%%%%%%%%%%%%%%%%%%%%%%%%%%%
	\begin{figure}%f11
	\centering
	\includegraphics[width = \columnwidth  ]{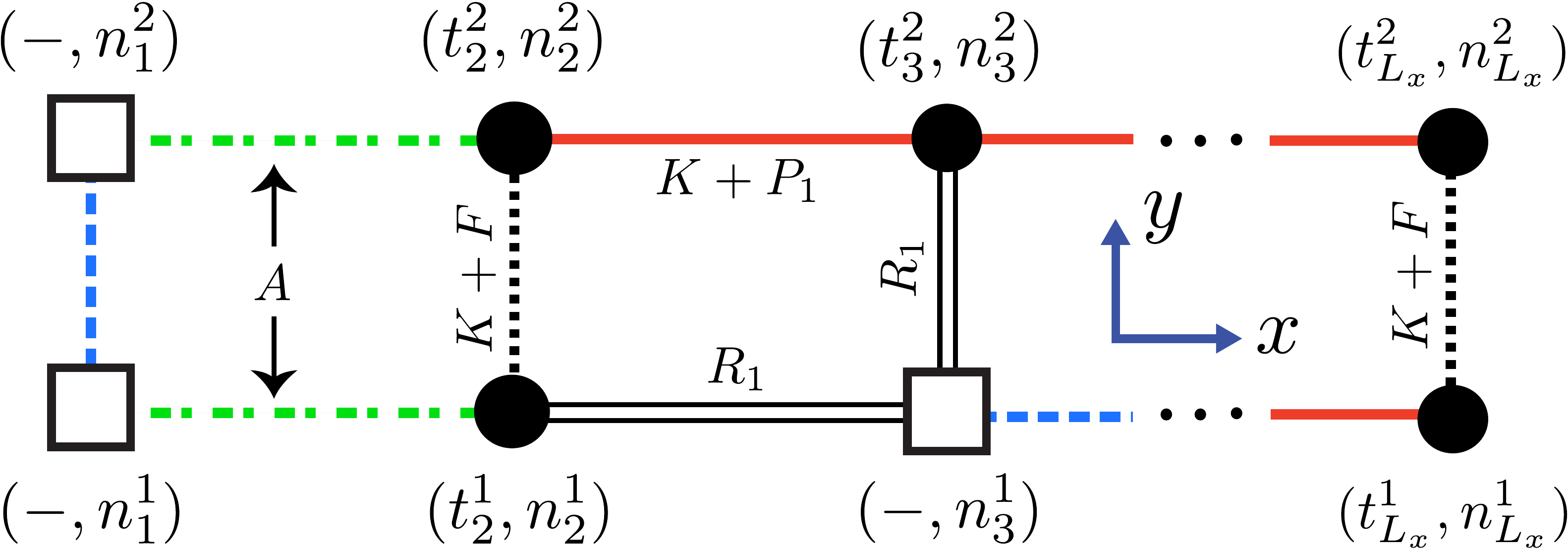}
	\caption{Proteins or solvent clusters may occupy lattice sites, where the front-view \protect ({\em y}-{\em z} plane) of an aggregate of A$\beta$(1-40) proteins is shown along with the interactions between proteins and solvent clusters. The $n_c=2$ nucleus is represented by dashed-dotted lines (free energy $A$ denoting the nucleation). Dotted and solid lines illustrate interactions between sheet proteins. Double solid lines illustrate a protein-solvent interface. Dashed (blue) lines have no meaning. }
	\label{ab40dilute}
	\end{figure}
%%%%%%%%%%%%%%%%%%%%%%%%%%%%%%%%%%%%%%%%

For either description of fibrils (model A or B), the total number of proteins on a strip lattice is then $N_{strip}\equiv\sum_{i=1}^{N_T} \sum_{j=1}^{L_y} n_{i}^{j}$. The grand partition function can be written as: 
\begin{equation}
\label{strip_pf}
\mathcal{Q}_{strip}^{A(B)} = \sum_{\{t\}, \{n\}} \exp\left(-\beta \mathcal{H}_{strip}^{A(B)} + \beta \mu_{PC} N_{strip}\right) 
\end{equation}
where the sums over ${\{t\}, \{n\}}$ are for all $i$ and $j$, and A, B refers to the effective Hamiltonians given by Equation~(\ref{two_filament}) or (\ref{hamilb}), respectively. For periodic boundary conditions, Equation~(\ref{strip_pf}) can be solved as $\mathcal{Q}_{strip}^{A(B)}=\operatorname{tr} \left(T_{strip}^{A(B)} \right)^N$ where $T_{strip}^{A(B)}$ is the partition function for the lattice-gas model (A or B). Just as in Subsection~3.1, in the thermodynamic limit $N_T \to\infty$:
\begin{equation}
\label{strippf}
\left(L_y N_{T}\right)^{-1} \ln \mathcal{Q}_{\text{strip}}^{A(B)} = \ln \lambda_{1}^{A(B)}
\end{equation} 
where $\lambda_{1}^{A(B)}$ is the largest eigenvalue of $T_{strip}^{A(B)}$. In general, the dimension of the transfer matrix $T_{strip}^{A}$ is $(q+1)^{n_c L_y} \times (q+1)^{n_c L_y}$ and has $(q+1)^{n_c L_y}$ number of eigenvalues, whereas the transfer matrix $T_{strip}^{B}$ is $(q+1)^{L_y} \times (q+1)^{L_y}$ and has $(q+1)^{L_y}$ number of eigenvalues.

To compare with experiments, we can define quantities similar to Equations~(\ref{thetaks}) and (\ref{aveleng}), and others. For example, in the grand canonical ensemble, the average number of proteins on the lattice, $\langle N_p \rangle$, referred to as the occupation of the lattice, the number of proteins in filaments, $\langle \psi \rangle$, the average number of filaments, $\langle \gamma \rangle$, and the average number of sheet segments, $\langle \theta \rangle$, can be written as: 
\begin{eqnarray}
\label{therm1}
\langle N_{p} \rangle &\equiv& z \frac{\partial}{\partial z} \ln \mathcal{Q} \\ 
\label{therm2} 
\langle \psi \rangle  &\equiv& \frac{\partial}{\partial K} \ln \mathcal{Q} + \langle \gamma \rangle  \\ 
\label{therm5}
\langle \theta \rangle &\equiv&  \frac{\partial}{\partial P_1} \ln \mathcal{Q} \\
\langle \gamma \rangle &\equiv&  \frac{1}{2} \frac{\partial}{\partial A} \ln \mathcal{Q} 
\end{eqnarray}
respectively. Other quantities may also be defined including the average lengths of filaments, and the average length of sheet stretches in aggregates~\cite{Schreck2011, Hong2011}. 

\section{Comparison to Experiment}
%%%%%%%%%%%%%%%%%%%%%%%%%%%%%%%%%%%%%%%%
	\begin{figure*}%[t]
	\begin{center}
	\includegraphics[width = 1.7\columnwidth]{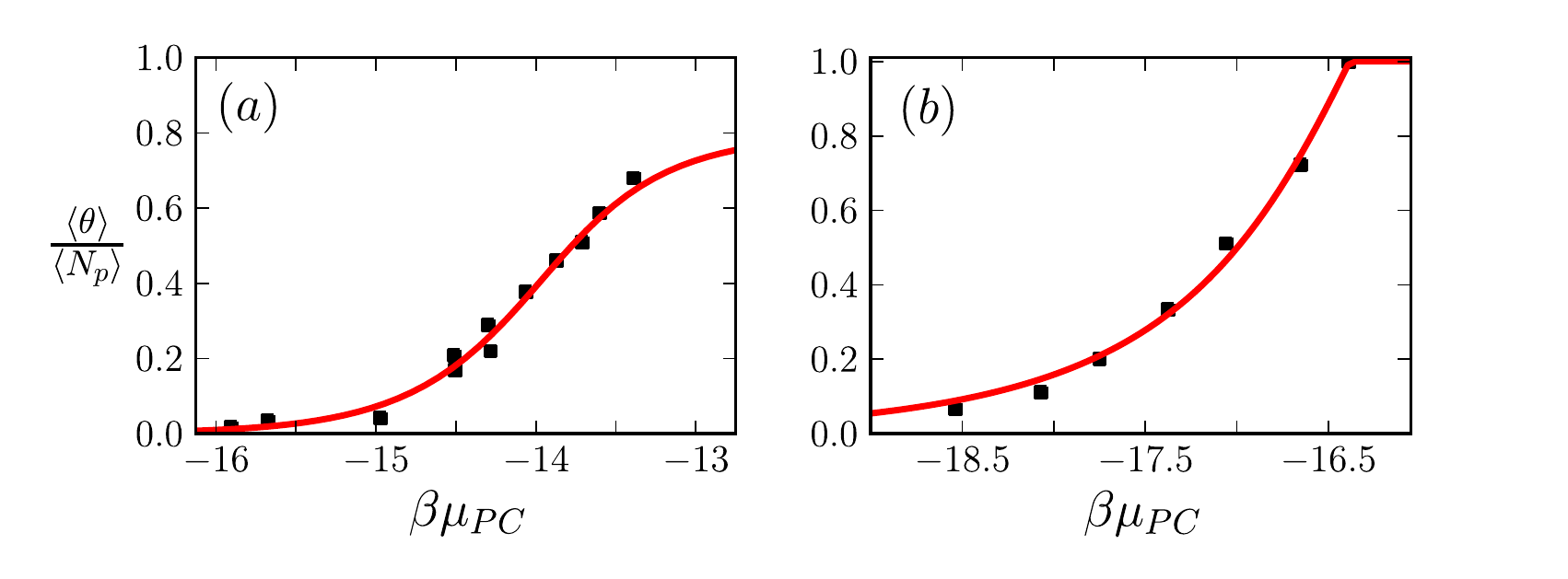}
	\caption{ ({\bf a}) $\langle \theta \rangle$/$\langle N_p \rangle$ is fitted to the results of the Terzi {\em et al.} experiment~\cite{terzi} involving $A\beta$(1-40) aggregates; ({\bf b}) The fraction of sheet proteins in Curli fibrils is fitted to the scaled results of the Hammer {\em et al.} experiment~\cite{hammer}. In ({\bf a}), the fit parameters were $P_1 \approx K \approx~A \approx~0$ kcal/mol, $R_1=0.35$ kcal/mol, and $F=16.4$ kcal/mol.; while in ({\bf b}) we have $P_1=7.26$ kcal/mol, $K=2.2$ kcal/mol, $R_1\approx0$ kcal/mol, and $A=1.2$~kcal/mol~\cite{Schreck2011}. In ({\bf a}) we used case B of the strip models with $n_{c}=2$ whereas in ({\bf b}) we used the 1D model with $n_c = 2$ for aggregation. In both cases $q=2$.} 
	\label{fits}
	\end{center}
	\end{figure*}
%%%%%%%%%%%%%%%%%%%%%%%%%%%%%%%%%%%%%%%%
\label{experimental}
We solve for $\mu_{PC}$ for $A\beta$(1-40) from Equation~(\ref{equib}) in terms of $\mu_{ST}$, $\mu_{SR}$, $\mu_{PV}$, and experimental concentration $c$. Then, $\mu_{PC}$ is plugged into Equation~(\ref{strip_pf}), and Equation~(\ref{therm5}) and other thermodynamic quantities can be calculated. For A$\beta$(1-40), we have $\mu_{ST}+\mu_{SR}\approx -29$ kcal/mol~\cite{Ferrone2006, Hill1987}. In reference~\cite{Ferrone2006}, $\mu_{PV}$ for hemoglobin was found to be approximately $\frac{3}{4}$($\mu_{ST}$+$\mu_{SR}$). Mutated hemoglobin does polymerize as amyloid, but amyloid proteins usually may be natively unstructured, unlike hemoglobin. We nevertheless use a similar result for $\mu_{PV}$ for $A\beta$(1-40) and Curli. Equation~(\ref{therm5}) divided by Equation~(\ref{therm1}), $\langle \theta \rangle / \langle N_p \rangle$, the $\beta$-sheet fraction, is used as our fitting function. The results are plotted in Figure~\ref{fits}a for $A\beta$(1-40)  fibrils, and Figure~\ref{fits}b for the Curli fibrils. The fit yields reasonable free energies at room temperature for the A$\beta$(1-40) fibrils, {$P_1 \approx K \approx$ $A \approx$  0 kcal/mol, $R_1=0.35$ kcal/mol, and $F=16.4$ kcal/mol. For the Curli fibrils, we found  $P_1=7.26$ kcal/mol, $K=2.2$ kcal/mol, $R_1\approx0$ kcal/mol, and $A=1.2$ kcal/mol~\cite{Schreck2011}. Clearly for the experiment involving A$\beta$(1-40) fibrils, the grand canonical approach to modeling fibrils does a better job than earlier canonical approaches. The grand-canonical model also suggests that the A$\beta$ fibrils are more strongly held together by inter-filament interactions when compared to the fit from the canonical model, and also that the penalty in going from sheet to coil regions in the aggregates is very small and could be due to modeling uncertainty. The minimum number of parameters needed to fit the CD data is also the same number when compared with the van Raaij and Schmidt's models, and if the value of $R_1$ in the A$\beta$ fit is taken to be within modeling uncertainty, then only one parameter is needed to fit the CD~data.

\section{Conclusions}

By focusing on the aggregation of proteins in forming oligomers, protofibrils,
and fibrils, and their relations to neurodegenerative diseases, statistical mechanical
approaches to protein aggregation have been developed. We have made a general summary
of the field, presenting recent formulations of the ZB model based on the canonical, as well as
the grand canonical, approaches to the amyloid formation processes. Some results
are presented to show that these models can be used to interpret experimental
observations as well as to provide phase diagrams~\cite{Schreck2011} showing the parameter dependence of
the $\beta$-sheet dominating regions. 

More experimental data like the CD results of Terzi {\em et al.}~\cite{terzi} and the AFM measurements performed by van Raaij {\em et al.}~\cite{dutch_2008} would help validate the ZB approach. For example, the ZB model for protein folding has been used to classify all the amino acids based on their propensity to fold from coil to helix~\cite{Scheraga2002}. Similar classification schemes could potentially be devised for the many proteins that can aggregate to form fibrils if a much larger collection of experimental data (like the CD and AFM results) were available. 

Of course, a statistical mechanical approach to protein aggregation has serious limitations. It can only be used to study the equilibrium properties of the systems, not the rates of the processes involved nor the transient behaviors, such as quasi-equilibrium or kinetic trapping. Furthermore, the available experimental data that we can compare our theories with are so far extremely limited. Therefore, statistical mechanical models are not a tool for predicting assembly pathways. However, statistical mechanics can be used to show that experimental observations are consistent with the predictions of a certain route of aggregation, that is, given a route of aggregation and the associated effective free energy, statistical mechanical models can predict equilibrium distributions of oligomers, protofibrils, fibrils, {\em etc}., which can then be compared to observed data.

The pathway prediction function is better achieved by using kinetic models~\cite{Knowles2009, Pappu2011, Ricchiuto2012, Hong2013, Schreck2013}, or better still, by molecular dynamics simulations~\cite{thirumalai2003, peng2004discrete, Ma2006, hall2006simulations, straub2011, morriss2012}. Unfortunately, the latter are highly restricted by the
system size and length of time that molecular dynamics can be used to simulate. On the other hand,
protein aggregation, as discussed earlier, is a complex process spanning many levels of
structure and many chemical species. Thus, at present, the kinetic models or coarse-graining models may be better tools for the purpose of predicting pathways. Moreover, many more experimental data are
available in the literature for  non-equilibrium or kinetic studies. It would be interesting, for example, to
develop a kinetic approach based on statistical mechanics, similar to a
kinetic Ising model derived from an Ising model. The kinetic study of protein aggregation is a
rich field of investigation and of great current interest. We hope to be able to report progress in the non-equilibrium studies of protein aggregation in future work. 

\section*{\noindent Acknowledgements}
\vspace{12pt}

We would like to thank Drs Frank Ferrone, Brigita Urbanc, and J. van Gestel for stimulating discussions.

\section*{\noindent  Conflicts of Interest} 
\vspace{12pt}

The authors declare no conflicts of interest.

%==========================================================
%==========================================================
% Back Matter (References and Notes)
%----------------------------------------------------------
% Style and layout of the references
\bibliographystyle{mdpi}
\makeatletter
\renewcommand\@biblabel[1]{#1. }
\makeatother
%----------------------------------------------------------
%\bibliography{meta}

\end{document}